\documentclass{optica-article}

\journal{opticajournal} 

\articletype{Research Article}
\usepackage{comment}
\usepackage{lineno}
\usepackage{cleveref}
\crefname{equation}{Eq.}{Eq.} 
\crefrangelabelformat{equation}{(#3#1#4)--(#5#2#6)}
\crefname{figure}{Fig.}{Fig.}

\usepackage{graphicx}
\usepackage{subcaption}
\usepackage{multirow}
\usepackage{floatrow}
\usepackage{amsmath}
\usepackage[utf8]{inputenc}
\usepackage[export]{adjustbox}
\begin{document}

\title{Effects of cavity nonlinearities and linear losses on silicon microring-based reservoir computing}

\author{Bernard J. Giron Castro\authormark{1*}, Christophe Peucheret\authormark{2}, Darko Zibar\authormark{1} and Francesco Da Ros\authormark{1}}

\address{\authormark{1} Department of Electrical and Photonics Engineering, Technical University of Denmark (DTU), 343 Ørsteds Plads, DK-2800 Kongens Lyngby, Denmark\\
\authormark{2}Univ Rennes, CNRS, UMR6082 - FOTON, 22305 Lannion, France\\}

\email{\authormark{*}bjgca@dtu.dk} 


\begin{abstract*} 
Microring resonators (MRRs) are promising devices for time-delay photonic reservoir computing, but the impact of the different physical effects taking place in the MRRs on the reservoir computing performance is yet to be fully understood. We numerically analyze the impact of linear losses as well as thermo-optic and free-carrier effects relaxation times on the prediction error of the time-series task NARMA-10. We demonstrate the existence of three regions, defined by the input power and the frequency detuning between the optical source and the microring resonance, that reveal the cavity transition from linear to nonlinear regimes. One of these regions offers very low error in time-series prediction under relatively low input power and number of nodes while the other regions either lack nonlinearity or become unstable. This study provides insight into the design of the MRR and the optimization of its physical properties for improving the prediction performance of time-delay reservoir computing.

\end{abstract*}

\section{Introduction}
Neuromorphic computing systems, which try to resemble the working mechanism of the human brain, are an interesting alternative to traditional Von Neumann architectures. Fundamental physical boundaries of electronics set some limits in future developments of current architectures to increase their computing capacity. Hence, neuromorphic computing appears to be a promising step in the development of novel artificial intelligence processors that can enhance the performance of current computing architectures and might extend Moore's law \cite{Yadav2023}. Over the last decade, developments in integrated photonics have allowed the exploration of novel computing paradigms. Additionally, these developments have driven the photonic hardware realization of computing processing schemes and machine learning algorithms. Lower energy consumption, parallel computing and faster processing speed are the key potential benefits of photonic computing architectures that could address the limitations of traditional electronic circuits 
\cite{Dabos:22, s22030720}. Photonic neural networks, all-optical switching, optical spiking neurons and optical activation functions are some examples of the emergence of the photonic computing field \cite{s22030720}. 

Reservoir computing (RC) is a relatively recent computing paradigm in the recurrent neural networks (RNNs) family that offers a lower complexity of the training process with respect to conventional RNN and other neural network schemes \cite{Appeltant2011InformationPU, LUKOSEVICIUS2009127 }. An RC architecture consists of an input layer, in which the data is assigned random fixed weights before being transferred to the reservoir layer, where the data is mapped into a higher dimensional space by means of interconnected nonlinear nodes with random and fixed connections. Using the response of the reservoir nodes, the weights in the output layer are trained to solve a specific target task, usually using ridge or linear regression. Based on the trained weights, RC can also make a prediction of the target task for subsequent input sequences that are unknown to the reservoir. Only the output layer is trained in RC, and this key feature considerably decreases the training time of this type of neural network \cite{LUKOSEVICIUS2009127, LUKOSEVICIUS2012}. RC has been shown to have applications in time-series predictions, channel equalization, speech recognition, medical and financial applications, etc. Further details about RC architecture are available in comprehensive reviews on the subject \cite{ LUKOSEVICIUS2009127, LUKOSEVICIUS2012}.

Multiple works have demonstrated implementations of photonic RC, where usually the nonlinear nodes are achieved through the nonlinear behaviour of photonic devices \cite{Vandoorne:08, Duport:12, Larger:12, Paquot2012, Mesaritakis:13, Vandoorne2014, Bueno:17, Chen:19, Vatin:19, Borghi2021, Nakajima2021, Donati:22, Abdalla:23, WANG} or the dynamics of nonlinear optical phenomena \cite{Mesaritakis:19, 8807158, Bu:22}. Some of these works consist of blocks of photonic devices that perform as nonlinear nodes \cite{Vandoorne:08, Mesaritakis:13, Vandoorne2014}, but this leads to the scalability of RC being a challenge as well as the footprint of the photonic circuit being considerably increased. An alternative approach known as time-delay reservoir computing (TDRC) is to multiplex the nodes in time and use a single physical nonlinear node that typically receives feedback through a physical loop to boost the connectivity between the virtual nodes and the overall memory of RC. Several works regarding photonic TDRC can be found in the literature, e.g., using a Mach Zehnder modulator as the nonlinear node \cite{Larger:12, Paquot2012, Chen:19, Nakajima2021, Abdalla:23}, semiconductor optical amplifiers \cite{Duport:12} or the nonlinear dynamics of laser devices \cite{Bueno:17, Vatin:19}. A TDRC setup based on microring resonators (MRR), first studied in \cite{Donati:22}, demonstrated a good performance in time-series prediction tasks. Nonetheless, there was no clearly established relationship between the performance of RC and the physical effects that generate the nonlinear dynamics of the microring cavity. In \cite{castro2023impact} we reported initial studies on the impact of varying the relaxation times of such physical effects and numerically showed that it is possible to obtain frequency detuning and power regions with a prediction error lower than other RC implementations with a similar number of virtual nodes and input rate. 

In this work, we extend our study of the MRR-based TDRC architecture from \cite{castro2023impact} by investigating the impact of the cavity waveguide linear loss on the performance of RC. This study also encompasses an analysis of how the amount of nonlinearity given by the cavity dynamics influences RC and how the behaviour of such dynamics can be used to improve the performance of this type of RC implementation. We also explore the impact of the generated nonlinear oscillations in the time-series prediction and show that it is a non-desirable effect for solving the discrete-time tenth-order nonlinear auto-regressive moving average (NARMA-10) task. Additionally, this work deepens the understanding of the performance thresholds and the fabrication requirements for the microring waveguide in order to achieve lower error prediction than similar numerical RC schemes. This improvement is shown for the prediction of the NARMA-10 sequence and can potentially be extended to other RC tasks.

The structure of the paper goes as follows: In section \ref{section2} we introduce the model used to mathematically describe the nonlinear dynamics of the MRR-based TDRC scheme. In this section, we also detail the major assumptions and values used for the optical parameters. In section \ref{section3} we present the details of each of the TDRC layers, including the mathematical description of the electric field of the processed optical signal at the different stages of our setup. In section \ref{section4} we describe the benchmark methodology when solving the NARMA-10 task. In section \ref{section5} we present the results of the setup when varying different physical parameters related to the MRR properties, and define input power vs. frequency detuning regions with different levels of prediction error. Afterwards, in section \ref{section6} we analyze the previous results and relate the region of low error prediction with the physical properties of the MRR and the dominance of each of the nonlinear effects. Section \ref{section7} summarizes the main conclusions.
\medskip
\section{Free-carrier nonlinearities in silicon MRR} \label{section2}

Silicon microring resonators have been extensively studied in the field of photonic computing as their features have demonstrated applications in a variety of computing processes such as all-optical switching \cite{6783786, Forst:07, Waldow:08}, optical logic gates \cite{Xiong:13}, weight banks \cite{Tait2017}, photonic spiking neural networks \cite{Xiang:22}, photonic accelerators \cite{Zhou2022} and photonic RC \cite{Borghi2021, Donati:22}. The study in \cite{Donati:22} focuses on the dependence of the MRR-based TDRC performance on the feedback intensity and the phase shift of the external feedback waveguide. This analysis is tested for the NARMA-10 and Mackey-Glass tasks. The dependence on the input power and detuning of the pump for the Santa Fe task is also studied. The system investigated in \cite{Borghi2021} presents a similar scheme but without external feedback and was tested on analog and binary operations. Furthermore, the impact of noise on the performance of the previous schemes with or without feedback was analyzed in \cite{2022SPIE12004E..0UD}.

To model the dynamics of a silicon MRR, we use the same mathematical model as in our previous study \cite{castro2023impact}, similar to the one used in \cite{Donati:22}. In this model, we introduce in the input port of the MRR a quasi-monochromatic electric field $E_{in}$ at an angular frequency $\omega_p$ close to that of the resonance frequency of the cold MRR cavity $\omega_0$. This field triggers the generation of excess carriers due to two-photon absorption (TPA). The conversion of the optical mode energy to heat by the absorption of power results in the increase of the cavity temperature. The generated free carriers change the refractive index of the cavity waveguide through free-carrier dispersion (FCD), which in turn results in a blue shift of the resonance frequency. They also become a source of free-carrier absorption (FCA). FCA contributes to the total rise of heat in the cavity. The resulting thermo-optic (TO) effect also changes the refractive index of the cavity but in the opposite direction, causing a red shift of the resonance \cite{Johnson:06}. All of these effects taking place inside the MRR cavity can be described by the temporal coupled mode theory (TCMT) through the following system of coupled differential equations, for an add-drop MRR configuration\cite{Donati:22, Johnson:06, Soltani2009NovelIS, VanVaerenbergh:12, Mancinelli2013LinearAN, PhysRevA.86.063808}: 
\begin{equation}\label{eq1}
\frac{\textrm da(t)}{\textrm dt} = [i\delta_\omega(t)-\gamma_{\textrm {tot}}(t)]a(t) + i
	\sqrt{\frac{2}{\tau_\textrm c}}\left[E_{\textrm {in}}(t) + E_{\textrm {add}}(t)\right]e^{i\omega_\textrm pt},
\end{equation}
\begin{equation}\label{eq2}
\frac{\textrm d\Delta N(t)}{\textrm dt} = -\frac{\Delta N(t)}{\tau_{\textrm {FC}}} + 
        \frac{\Gamma_{\textrm {FCA}}c^2 \beta_{\textrm {TPA}}}{2\hbar\omega_pV^2_{\textrm {FCA}}n^2_{\textrm {Si}}} |a(t)|^4,
\end{equation}
\begin{equation}\label{eq3}
\frac{\textrm d\Delta T(t)}{\textrm dt} = -\frac{\Delta T(t)}{\tau_{\textrm {th}}} + 
        \frac{\Gamma_{\textrm {th}}P_{\textrm {abs}}(t)}{mc_\textrm p} |a(t)|^2,
\end{equation}

\noindent where $a$ is the modal amplitude within the resonator cavity, $\Delta N$ represents the excess free-carrier density generated via TPA, and $\Delta T$ is the temperature difference of the waveguide cavity with respect to the environment. The variation of $a$ described in Eq. \ref{eq1} is dependent on both $E_{\textrm {in}}$ and $E_{\textrm {add}}$, where the latter denotes the electric field at the add port in an add-drop MRR configuration. 1/$\tau_\textrm c$ is the decay rate of the cavity modal energy due to the coupling of each bus waveguide to the MRR. The terms $\delta_\omega (t)$ and $\gamma_{\textrm {tot}}(t)$ represent the total angular frequency detuning and losses rate in the MRR cavity, respectively. In \cref{eq2,eq3}, $c$ and $\hbar$ denote the speed of light in vacuum and the reduced Planck's constant, respectively. $\tau_{\textrm {FC}}$ is the relaxation time of the free carriers, $\tau_{\textrm {th}}$ is the decay time of the TO effect, and $m$ is the mass of the MRR. The terms $\beta_{\textrm {TPA}}$, $n_{\textrm {Si}}$ and $c_{\textrm p}$ refer to the TPA coefficient, refractive index and specific heat of silicon, respectively. $\Gamma_{\textrm {FCA}}$ and $\Gamma_{\textrm {th}}$ are the FCA and thermal confinement factors related to the fractional energy overlap of the mode with the differential temperature and excess of FCD within the silicon microring. $V_{\textrm {FCA}}$ is the effective volume of the FCA. $P_{\textrm {abs}}(t)$ is the total optical mode energy converted into absorbed power. The time-dependent terms $\delta_\omega (t)$, $\gamma_{\textrm {tot}}(t)$ and $P_{\textrm {abs}}(t)$ have the following definitions \cite{Johnson:06, VanVaerenbergh:12, Mancinelli2013LinearAN}:

\begin{equation}\label{eq4}
    \delta_\omega(t) = \omega_\textrm p - \omega_0\left[1 - \frac{1}{n_{\textrm {Si}}}\left(\Delta N(t) 
     \frac{\textrm dn_{\textrm {Si}}}{\textrm dN} + \Delta T(t) \frac{\textrm dn_{\textrm {Si}}}{\textrm dT} \right)\right],
\end{equation}

\begin{equation}\label{eq5}
    \gamma_{\textrm {tot}}(t) = \frac{c\alpha}{n_{\textrm {Si}}} + \frac{2}{\tau_\textrm c} + \gamma_{\textrm {TPA}} + \gamma_{\textrm {FCA}} = \frac{c\alpha}{n_{\textrm {Si}}} + \frac{2}{\tau_\textrm c} + \frac{\beta_{\textrm {TPA}}c^{2}}{n_{\textrm {Si}}^2V_{\textrm {TPA}}}|a(t)|^2 +\frac{\Gamma_{\textrm {FCA}}\sigma_{\textrm {FCA}}c}{2n_{Si}} \cdot \Delta N(t),
\end{equation}

\begin{equation}\label{eq6}
    P_{\textrm {abs}}(t) = \left(\frac{c\alpha}{n_{\textrm {Si}}} + \frac{\beta_{\textrm {TPA}}c^{2}}{n_{\textrm {Si}}^2V_{\textrm {TPA}}} |a(t)|^2 +\frac{\Gamma_{\textrm {FCA}}\sigma_{\textrm {FCA}}c}{2n_{Si}} \cdot \Delta N(t)\right)|a(t)|^2.
\end{equation}

In Eq. \ref{eq4}, the total angular frequency detuning is a result of the sum of the detuning between $\omega_\textrm p$ and $\omega_0$, which we refer to in this work as $\Delta\omega = \omega_\textrm p - \omega_0$, and the nonlinear detuning due to TO and FCD effects. $\textrm dn_{\textrm {Si}}/$$\textrm dT$ and $\textrm dn_{\textrm {Si}}/$$\textrm dN$ represent the TO and FCD coefficients of silicon. In \cref{eq5,eq6}, the terms $\gamma_{\textrm {TPA}}$ and $\gamma_{\textrm {FCA}}$ denote the losses due to TPA and FCA \cite{Johnson:06}, and $\alpha$ is the linear attenuation of the waveguide. $\sigma_{\textrm {FCA}}$ is the total FCA cross-section. The values of the optical parameters used for the simulation of the photonic RC are listed in Table \ref{tab1}.

\begin{table}[t!]
    \centering
    \begin{tabular}[c]{|c|c|c|c|} 
         \hline
         Parameter & Value & Parameter & Value\\
         \hline
         $m$   & $1.2\times10^{-11}$ kg    &$\beta_{\textrm {TPA}}$&   $8.4\times10^{-11}$ m $\cdot$ W$^{-1}$ \cite{VanVaerenbergh:12} \\
         $\tau_{\textrm c}$&   $54.7$ ps  & $\Gamma_{\textrm {FCA}}$  & 0.9996 \cite{VanVaerenbergh:12} \\
         $n_{\textrm {Si}}$ & 3.485 \cite{Johnson:06} & $\Gamma_{\textrm {th}}$  & 0.9355 \cite{VanVaerenbergh:12}\\
         $\lambda_{0} $ & $1553.49$ nm & d$n_{\textrm {Si}}/$d$T$ &  $1.86$ $\times$ 10$^{-4}$ K$^{-1}$ \cite{Johnson:06}\\
         $L$&   $2\pi\cdot7.5$ $\mu$m & d$n_{\textrm {Si}}/$d$N$ & $-1.73$ $\times$ 10$^{-27}$ m$^{-3}$ \cite{VanVaerenbergh:12}\\
         $c_{\textrm p}$& 0.7 J $\cdot$ (g $\cdot$ K)$^{-1} $ \cite{Johnson:06} & $\sigma_{\textrm {FCA}}$& 1.0 $\times$ 10$^{-21}$ m$^2$ \cite{VanVaerenbergh:12}\\
         $V_{\textrm {FCA}}$& 2.36 $\mu$m$^{3}$ \cite{VanVaerenbergh:12}& $V_{\textrm {TPA}}$& 2.59 $\mu$m$^{3}$ \cite{VanVaerenbergh:12}\\
         \hline
     
    \end{tabular}\par
    \caption{Optical parameters used in the photonic RC simulations.}
    \label{tab1}
\end{table}

Our model does not consider the contribution of the frequency detuning due to Kerr effect in Eq. \ref{eq1} as the induced refractive index change (of the cavity waveguide) is negligible compared to the change caused by TO and FCD effects \cite{PhysRevA.87.053805} at the simulated input power range. However, it is important to point out that the Kerr effect becomes more relevant in the case of coupled cavities and different material platforms, as recently studied in \cite{Boikov_2023}. The model also does not consider any source of noise, nor does it consider the counterpropagating optical mode as the absence of backscattering is assumed. This assumption follows the approaches of \cite{Soltani2009NovelIS, VanVaerenbergh:12, Mancinelli2013LinearAN, PhysRevA.86.063808, PhysRevA.87.053805}. In order to apply the model to our RC tasks, we normalize and solve \cref{eq1,eq2,eq3} using a $4^{\textrm{th}}$-order Runge-Kutta method similar to the solutions of the TCMT equations described in \cite{VanVaerenbergh:12, PhysRevA.87.053805}. Throughout this work, we sweep the values of the nonlinear effects lifetimes $\tau_{\textrm {FC}}$, and $\tau_{\textrm {th}}$, as well as the attenuation value given by $\alpha$. Then, we evaluate the impact of varying those parameters on RC dynamics and performance.

\section{MRR-based Time-delay photonic RC}\label{section3}

\begin{figure}[t!]
\begin{subfigure}{1.0\textwidth}
\centering\includegraphics[width=1.0\linewidth, trim={0 2.5cm 0 3.5cm}]{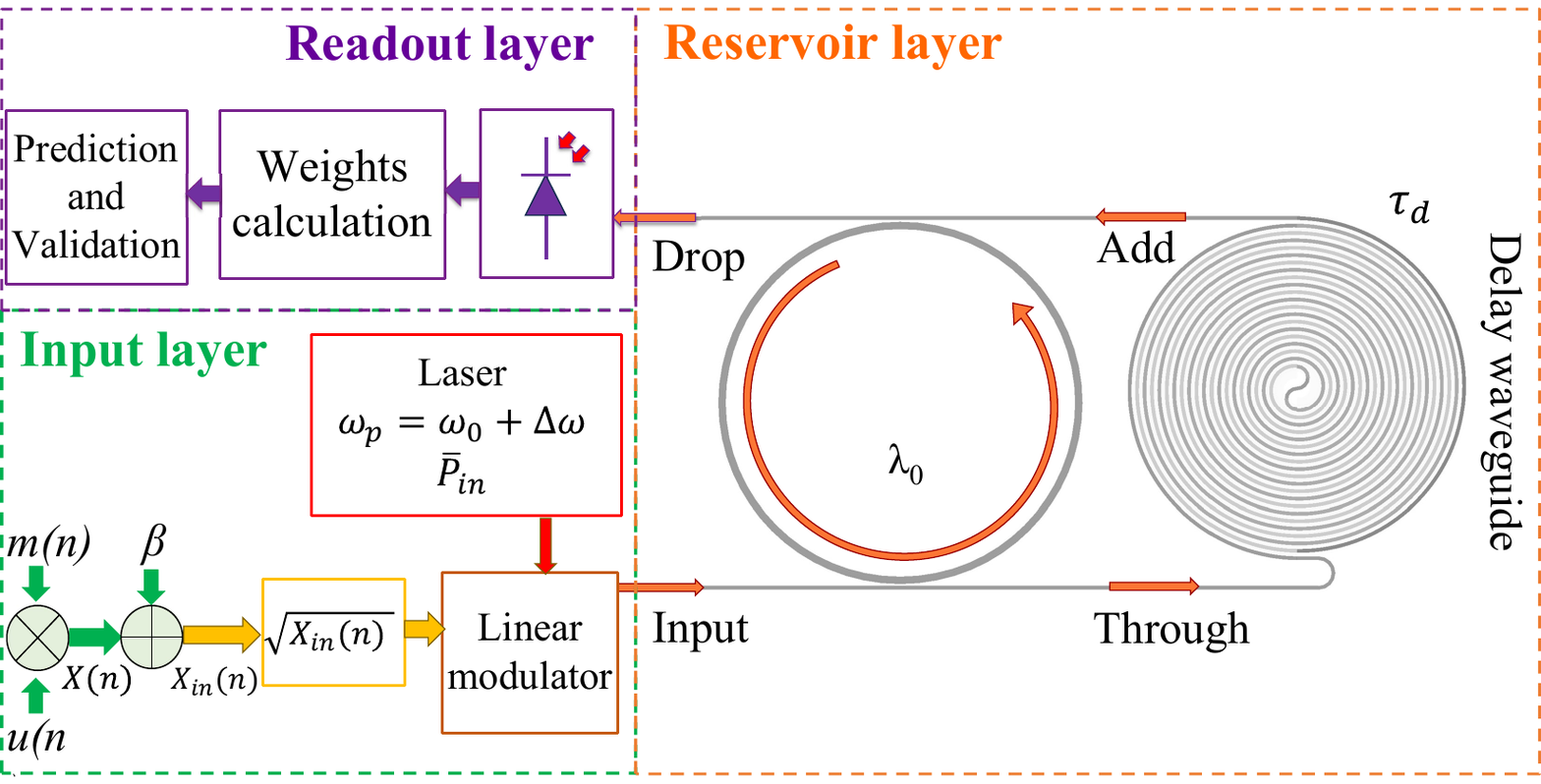} 
\caption{}
\label{fig:subim1}
\end{subfigure}

\begin{subfigure}{0.5\textwidth}
\includegraphics[width=1.0\linewidth, left]{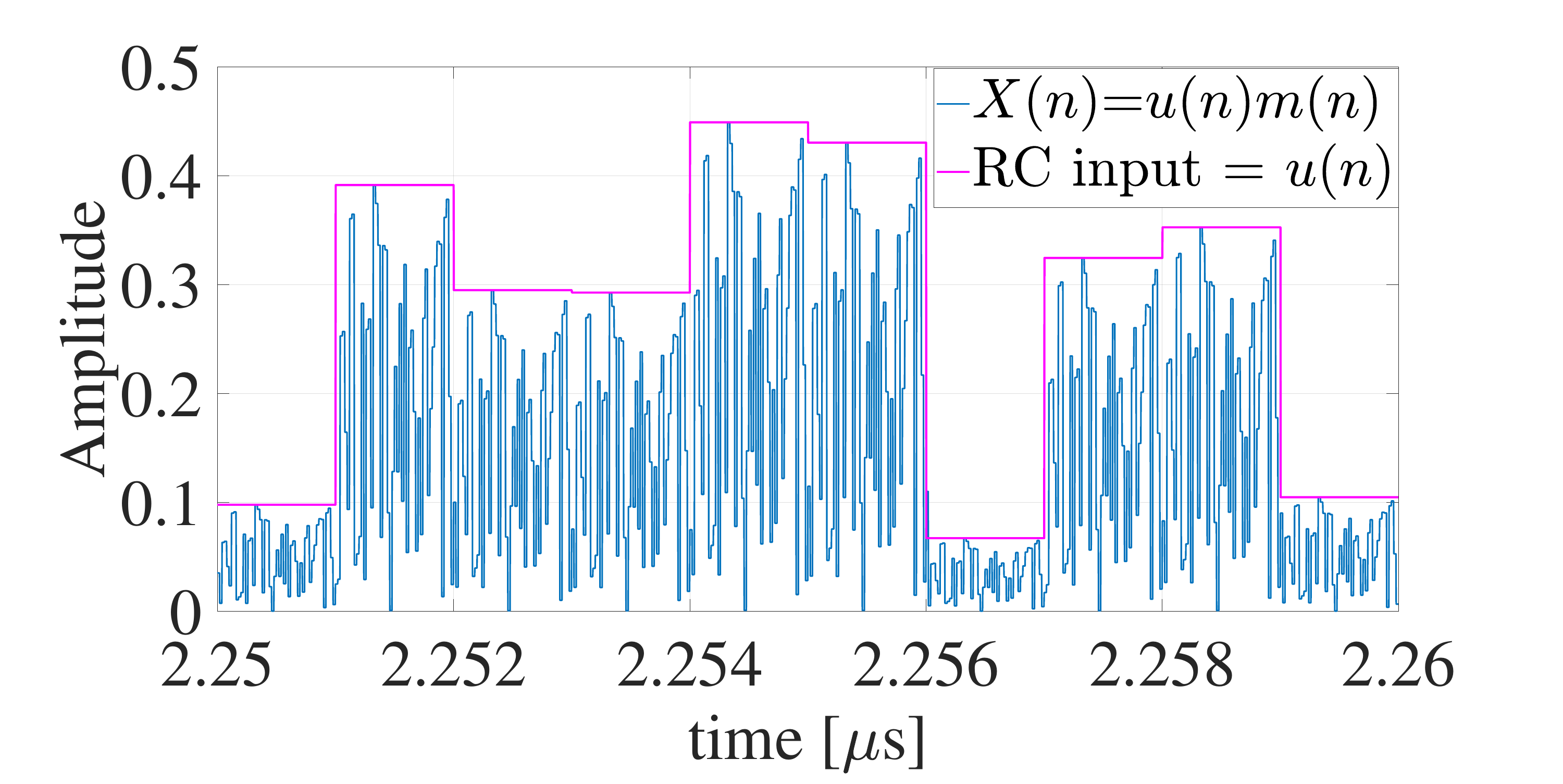}
\caption{}
\label{fig:subim2}
\end{subfigure}\hfill%
\begin{subfigure}{0.5\textwidth}
\includegraphics[width=1.0\linewidth, right]{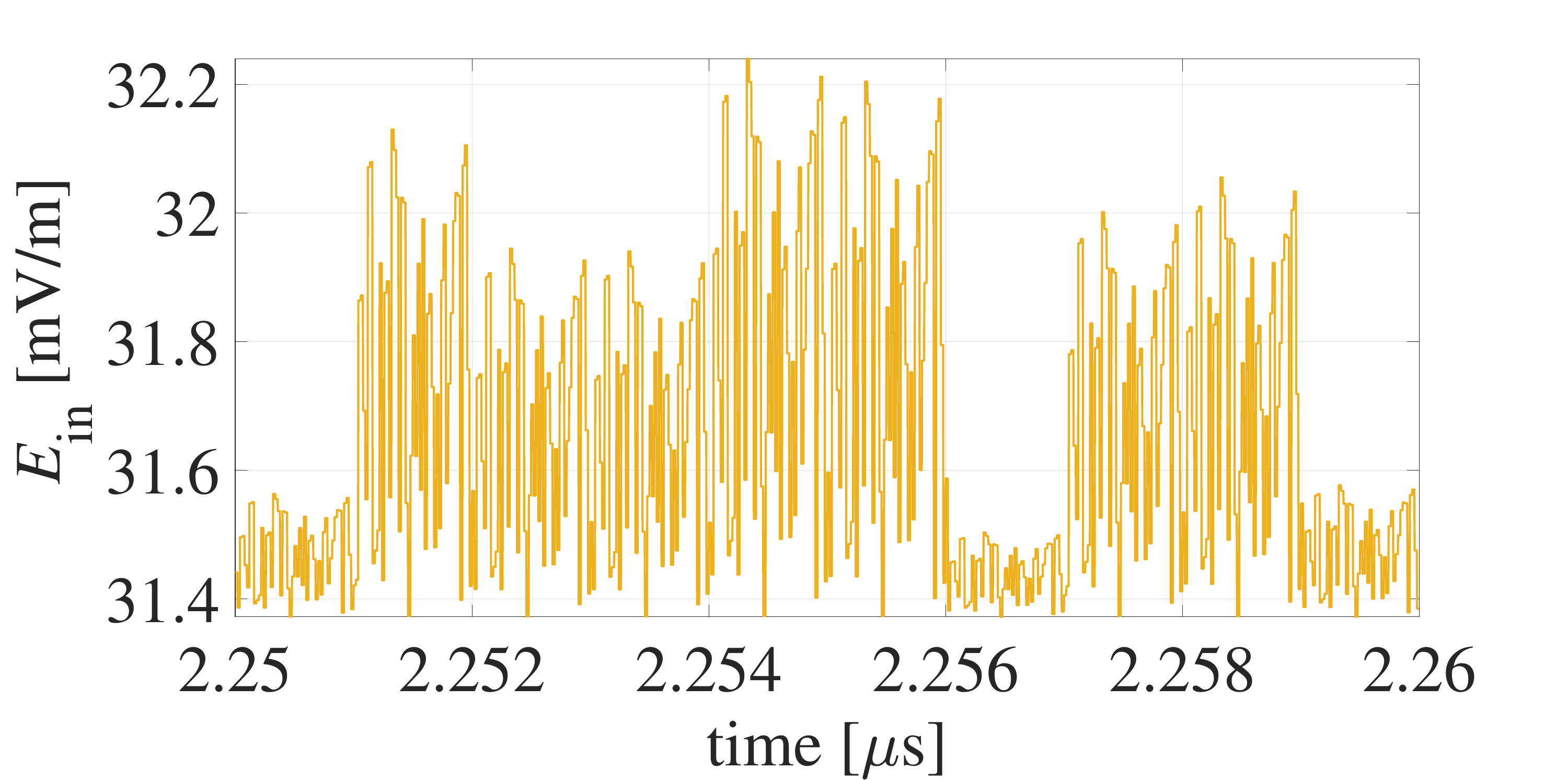}
\caption{}
\label{fig:subim3}
\end{subfigure}
\caption{a) Photonic TDRC scheme using a silicon MRR with delayed feedback. b) Data sequence and masked input of the RC, taken from an individual testing set of the RC simulations. c) Corresponding electric field envelope of the signal at the input port of the MRR. The small modulation index approximates a quasi-monochromatic optical signal.}
\label{fig:image1}
\end{figure}

The photonic TDRC simulated in this work (Fig. \ref{fig:subim1}), consists of an optical pump from a laser source which is modulated by the masked input data sequence of the RC. The virtual nodes are multiplexed in time by the masking signal $m(n)$ and a delay waveguide is added in order to increase the memory to the RC, with a length that matches the added delay. In \cite{Borghi2021,Donati:22}, the authors demonstrated the memory capacity provided by the MRR cavity itself and by the external waveguide. The response of the RC is obtained through a photodetector connected to the drop port of the MRR. Afterwards, the training and testing of the output layer are performed using ridge regression. 

The silicon MRR cavity acts as the single physical node of RC when nonlinear behaviour is induced. Indeed, this behaviour tends to become oscillatory as the FCD and TO effects cause an increase in cavity losses, which decreases the modal energy $|a|^2$. This reduction of energy in turn diminishes the nonlinear effects together with their caused losses, and consequently, $|a|^2$ starts to rise again, forming a cyclic nonlinear behaviour known as self-pulsing (SP) \cite{Soltani2009NovelIS, PhysRevA.86.063808, PhysRevA.87.053805}. 

In RC, the high nonlinearity is correlated with the dimensionality expansion that is required for computation tasks when their solution is not feasible in low dimensional input space, and a task-dependent correlation between higher dimensionality and better performance has been demonstrated \cite{Skalli:22, kärkkäinen2022dimensional}. 
However, a higher level of nonlinearity and dimensionality does not always entail a better performance of RC \cite{10.1063/1.5128898}, whereas, in time-series prediction tasks like the one analyzed in this work, it might even be detrimental for the memory capacity of the reservoir \cite{Inubushi2017}. Furthermore, in certain frequency detuning and input power conditions, a perturbation of the input optical signal can trigger self-pulsing of the cavity energy with ultra-short discontinuities or spikes, as previously studied on silicon MRR nonlinearities  \cite{PhysRevA.86.063808, Borghi:21}. Such fast pulse transitions can alter the stability of the RC dynamics, as they affect the computational consistency of the system, as defined in \cite{PhysRevLett.93.244102}. In \cite{Boikov_2023}, it is also pointed out that in the case of coupled cavities under the influence of FCD effects, there is an optimum input power interval in which the increase of dimensionality comes before the loss of consistency. As further discussed in section \ref{section5}, similar findings are achieved in this work where the RC reaches enough dimensionality to achieve good performance without exciting SP that could affect its stability. Further details from each of the layers of the simulated RC scheme are described in the following subsections.

\subsection{Input layer}
The 1-GBd input symbol sequence, $u(n)$ is multiplied by the mask $m(n)$, whose random values are generated from a uniform distribution over the interval [0, +1]. The size of $m(n)$ is determined by the number of nodes $N$ of the RC. Unless another value is specified, this work uses $N = 50$. Therefore, every symbol, which has a duration of 1.0 ns, belonging to the sequence $u(n)$,  is masked into a virtual node over a time interval of $\theta =\frac{1.0 \textrm{ ns}}{N} = 20$ ps, producing $X(n)$ (Fig. \ref{fig:subim2}). In order to fulfil the requirement of our model of using a quasi-monochromatic input electric field by using a small modulation index, we add a bias $\beta = 8.0$ to $X(n)$ which was optimized with respect to the performance of the RC (with the rest of the parameters fixed). The resulting masked input signal, $\hat{X}(n)$, has a modulation index of less than 2\%. $\hat{X}(n)$ is transformed into an electric signal that linearly modulates the optical field from an ideal noiseless continuous-wave laser source that generates an optical pump at a frequency $\omega_\textrm p$ and with average power $\overline{P}_{\textrm {in}}$. We denote the resulting linearly modulated optical signal power at the input port of the MRR as $X_{\textrm {in}}(n)$. Consequently, we can mathematically describe the input sequence $\hat{X}(n)$ and the electric field $E_{\textrm {in}}$ at the input port of the MRR (Fig. \ref{fig:subim3}) corresponding to the input optical signal (square root of $X_{\textrm {in}}(n)$), in the following way \cite{Donati:22}: 

\begin{equation}
    \hat{X}(n) = u(n)m(n) + \beta = X(n) + \beta,
\end{equation}

\begin{equation}
    E_{\textrm {in}}(n) = [X_{\textrm {in}}(n)]^{(1/2)}.
\end{equation}

\subsection{Reservoir layer}

The Runge-Kutta solution of \cref{eq1,eq2,eq3} is obtained with a step $\eta = 2.0$ ps. This value was sufficiently small to obtain an accurate solution for the cavity nonlinearities of the system under investigation in \cite{castro2023impact}. There are $M = \frac{\theta}{\eta} = 10$ solver steps per virtual node (500 per symbol for $N$ = 50). Hereafter, we sample and hold $E_{\textrm {in}}(n)$ for each $k^{\textrm{th}}$ step of the Runge-Kutta solver over an interval $\theta$ for each $j_{\textrm{th}}$ virtual node to build the input electric field $\hat{E}_{\textrm {in}}(k)$ used in our model solver :

\begin{equation}
    \hat{E}_{\textrm {in}}(k) =  E_{\textrm {in}}(n),  \qquad \textrm{for} \qquad (j-1)\theta \le (j-1)\theta + k\eta < j\theta, \qquad\qquad 0 \le k<M. 
\end{equation}

Similar to the mathematical description of the scheme performed in \cite{Donati:22}, the external waveguide provides a delay $\tau_\textrm d$ to the optical signal that links the through and the add port. Throughout the simulations of the RC, we use a value of $\tau_\textrm d = 0.5$ ns for the delay waveguide. This value was optimized as a function of the ratio between $\tau_d$ and the symbol duration (1.0 ns) where a delay of a duration of half the symbol length was found to be the optimum value. The solver steps equivalent to the delay time are determined as $\hat{\tau}_\textrm d = \tau_\textrm d / \eta = 250$. We assume no counterpropagating modes in the microring cavity. Hence, the electric field samples at the add and drop ports can be expressed as:

\begin{equation}
    \hat{E}_{\textrm {add}}(k) = \kappa e^{-i\phi_\textrm d}\left[\hat{E}_{\textrm {in}}(k-\hat{\tau}_\textrm d )+\frac{1}{\tau_\textrm c}a(k-\hat{\tau}_\textrm d )\right],
\end{equation}

\begin{equation}
    \hat{E}_{\textrm {drop}}(k) = \frac{1}{\tau_\textrm c}a(k)\hat{E}_{\textrm {in}}(k)+\hat{E}_{\textrm {add}}(k), 
\end{equation}

 \noindent where $\kappa$ = 0.95 represents the optimized coupling factor between the delay waveguide and the bus waveguides of the MRR. As in the case of $\beta$, $\kappa$  was optimized with respect to the performance of the RC (lowest prediction error). The obtained value of $\kappa$ is also close to the one used similarly in \cite{Donati:22}. The model takes into account the phase shift $\phi_\textrm d$ due to propagation through the delay waveguide for the optical pump with wavelength $\lambda_ \textrm p$. In \cite{Donati:22}, an adjustable external phase shift ($\Delta\phi$) in the delay waveguide was used to improve performance. In this work, $\Delta\phi$ = 0 is used, for which similar performance is achieved than for other values of $\Delta\phi$ within a limited range of configurations of the setup. The results obtained in this work could be further improved by a more systematic optimization of $\Delta\phi$, which is out of the scope of this work. $\phi_\textrm d$ is defined as:

\begin{equation}
    \phi_\textrm d = \frac{2\pi \tau_\textrm d c}{\lambda_\textrm p}.
\end{equation}

\subsection{Readout layer}

We average the $M$ step values of the Runge-Kutta solution for $\hat{E}_{\textrm {drop}}(k)$ over the duration $\theta$ for each $j_{\textrm{th}}$ virtual node to obtain $E_{\textrm {drop}}(n)$:

\begin{equation}
    E_{\textrm {drop}}(n) = \frac{1}{M}\sum_{k=(j-1)M+1}^{jM} \hat{E}_{\textrm {drop}}(k)
\end{equation}

Next, we simulate the photodiode response of the RC by calculating the square of $E_{\textrm {drop}}(n)$.

\begin{equation}
    X_{\textrm {drop}}(n) = \lvert E_{\textrm {drop}}(n)\rvert^2
\end{equation}

Once the response of the RC is obtained, we calculate the $N$-size weight vector $W$ for the output layer of the RC by using ridge regression with an optimized regularization parameter $\Lambda = 10^{-9}$. The impact of varying $\Lambda$ for different values of the physical parameters considered in this work is not analyzed and we keep it constant. The $n$ elements of the predicted sequence $\hat{y}$(n) are then determined as follows:

\begin{equation}
    \hat{y}(n) = \sum_{j=1}^N W_jX_{n,j}(n)
\end{equation}

\section{Benchmark}\label{section4}

To investigate the impact of the relaxation times of the studied nonlinear effects and the waveguide attenuation we test the system when solving a chaotic time-series prediction task like NARMA-10. This task requires the high dimensionality given by the nonlinearities of the cavity and the photodiode response, in addition to memory capabilities as it requires a memory of 10 steps in the past to be solved. The NARMA-10 time-series target equation can be expressed as \cite{846741}:

\begin{equation}
    y(n) = 0.3y(n) + 0.05y(n)\left[\sum_{i=0}^9 y(n-i)\right] + 1.5u(n-9)u(n) + 0.1 
\end{equation}

For this task, the sequence $u\left(n\right)$ is taken from a uniform distribution over the interval: [0.0, 0.5]. We use 2000 data points for the training and 2000 new data points for the testing. The whole data batch is processed into the RC at once. Additional to the 4000 data points, we add 200 warm-up data points at the start of both the training and the testing sets. This allows the Runge-Kutta solution to settle and eliminates any memory dependency between the sequences. To measure the performance of the RC, we calculate the normalized mean square error (NMSE) of the predicted sets. This metric is an estimation of the prediction errors, and so the lower the error, the higher the performance of the RC. The NMSE between the predicted sequence $\hat{y}(n)$, and the target $y(n)$ with a standard deviation $\sigma_{y}^2$ and length $L_{\textrm {data}}$, can be expressed as:
    
\begin{equation}
    \textrm {NMSE} = \frac{1}{L_{\textrm {data}}}\frac{\sum_{n} \left( \hat{y}(n) - y(n) \right)^2}{\sigma_{y}^2}    
\end{equation}

\section{Results}\label{section5}
In the following subsections, all the results are obtained from simulations under a $\overline{P}_{\textrm {in}}$ range of -20 to +20 dBm and a $\Delta\omega/2\pi$ range of $\pm 300$ GHz. For every simulation with the same value of $N$, we use the same generated random mask sequence $m(n)$. The calculated NMSE of our simulation results corresponds to the testing set and it is averaged over 10 different seeds used to generate $u(n)$ and consequently, the NARMA-10 sequence. As previously mentioned, a normalization process is carried out when solving \cref{eq1,eq2,eq3}. However, for visualization purposes, we present the denormalized quantities of $\Delta N(t)$ and $\Delta T(t)$ in our results. When not otherwise specified, we use the following values for the analyzed parameters: $\tau_ \textrm{th}$ = 50 ns, $\tau_ \textrm{FC}$ = 10 ns, $\alpha$ = 0.2 dB/cm, $N$= 50. The value of each parameter analyzed was selected as a middle point within realistic ranges of previously reported values from studies in the literature that use MRRs for photonic computing applications \cite{Soltani2009NovelIS, VanVaerenbergh:12, Mancinelli2013LinearAN, PhysRevA.86.063808, PhysRevA.87.053805}. First, using the previous parameters as the initial conditions we obtain the results of subsection \ref{sub5.1}.

\subsection{$\overline{P}_{\textrm {in}}$ vs $\Delta\omega/2\pi$ regions of NMSE}\label{sub5.1}
\begin{figure}[h!]
\centering\includegraphics[width=11.3cm, trim={0 4.0cm 0 3.0cm}]{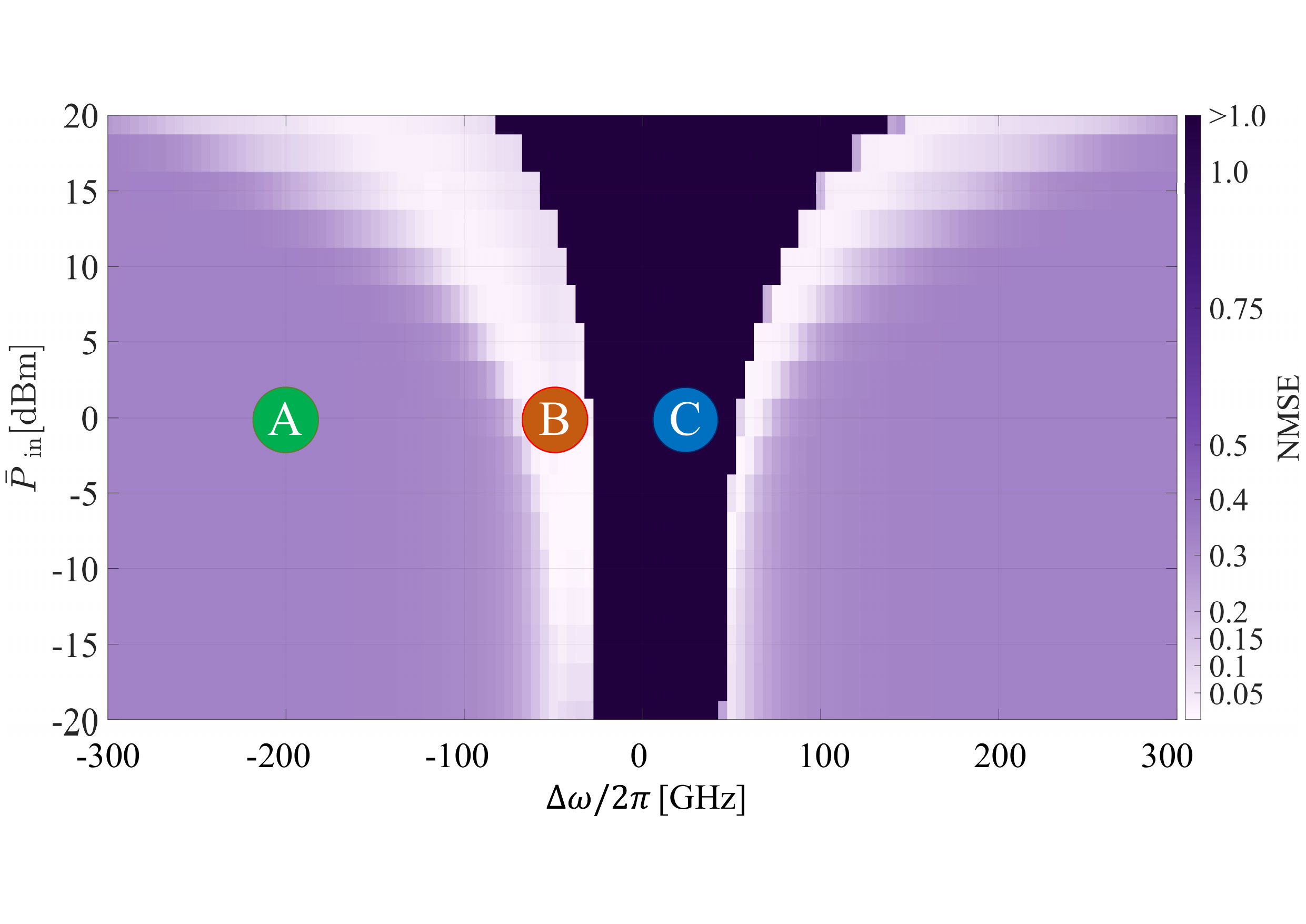}
\caption{The regions A, B, and C in terms of $\overline{P}_{\textrm {in}}$ vs $\Delta\omega/2\pi$ as defined in subsection \ref{sub5.1} when solving the NARMA-10 task ($N$ = 50, $\tau_ \textrm{th}$ = 50 ns,  $\tau_ \textrm{FC}$ = 10 ns, $\alpha$ = 0.8 dB/cm).}
\label{fig2}
\end{figure}

 We divide the parameter space $\overline{P}_{\textrm {in}}$ vs $\Delta\omega/2\pi$ according to the NMSE performance and identify three different behaviours. The corresponding regions are labelled A, B, and C, as shown in Fig. \ref{fig2}. The first region, A, occurs at higher detuning from the (cold) resonance of the MRR in the heat maps, in which the NMSE gradually approximates a constant value. The region C is located near the resonance of the MRR (although this is not necessarily the case for other parameter configurations, as further results demonstrate). It achieves an NMSE > 1.0, which indicates an inconsistent response of the RC which makes it unable to achieve accurate predictions. The third region, B is located in the middle of regions A and C at both the red-shift and blue-shift sides of the spectrum. This region is the one with the best performance of the RC (NMSE<0.2).

\begin{figure}[t!]
\centering\includegraphics[width=13.5cm, trim={1.75cm 1.5cm 1.75cm 2.5cm}]{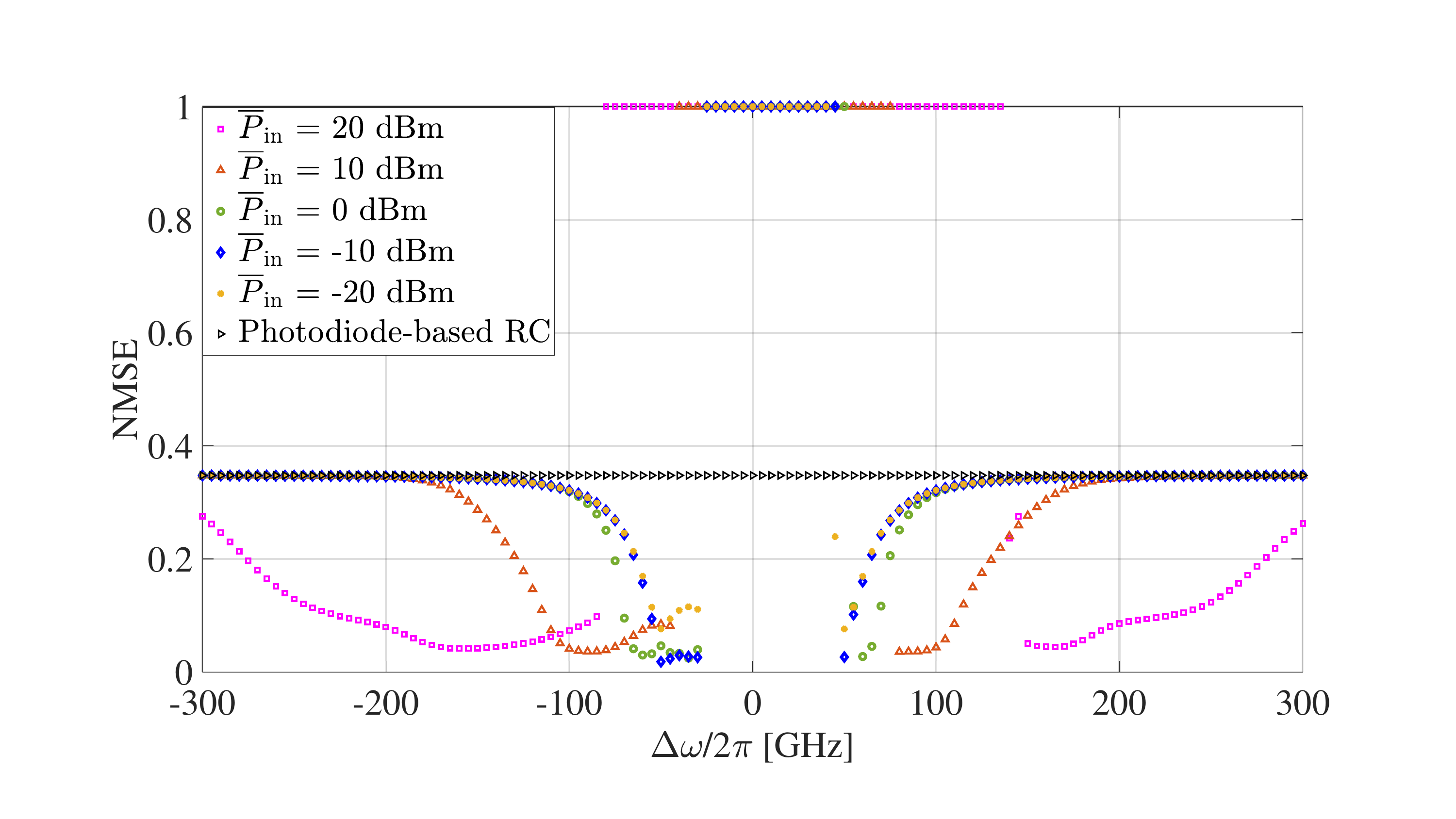}
\caption{NMSE as a function of $\Delta\omega$ for different levels of $\overline{P}_\textrm{in}$ ($\tau_ \textrm{FC}$ = 10 ns, $\tau_ \textrm{th}$ = 50 ns, $\alpha$ = 0.8 dB/cm). We also include the NMSE result obtained when the system uses a photodiode as single nonlinear element of the TDRC.}
\label{fig3}
\end{figure}

In Fig. \ref{fig3}, we show a slightly different perspective of the previous result by simulating a frequency sweep for a set of $\overline{P}_{\textrm {in}}$ values that highlight the transitions between the different regions. In order to determine the reasoning behind the constant value that region A appears to approximate when we increase either positive or negative detuning, we determine first the performance of the RC in the absence of nonlinear dynamics in the MRR. In this scenario, we can expect the photodiode response at the detection stage of Fig. \ref{fig:subim1} to become the only component acting as the physical nonlinear node of the RC, with the delay waveguide still present in the system. Therefore, we simulate our system using a photodiode as the nonlinear node and with absence of MRR nonlinearities, which we refer to as photodiode-based RC in this work. For such a setup, we obtain approximately the same NMSE threshold as the one to which region A converges in Fig. \ref{fig3} (NMSE $\approx$ 0.3475). This indicates that the MRR approaches a linear regime value the further we detune $\omega_p$ from $\omega_0$. The slope of this trend increases in low power levels and it decreases for 20 dBm, as the high power holds the FCD nonlinearities for a longer $\Delta\omega$ range. Later in this work, we also study the characteristic response from each of the regions. 

\subsection{Impact of the thermo-optic decay time}

\begin{figure}[h!]
\centering\includegraphics[width=13.3cm, trim={0 6.7cm 0 4.0cm}]{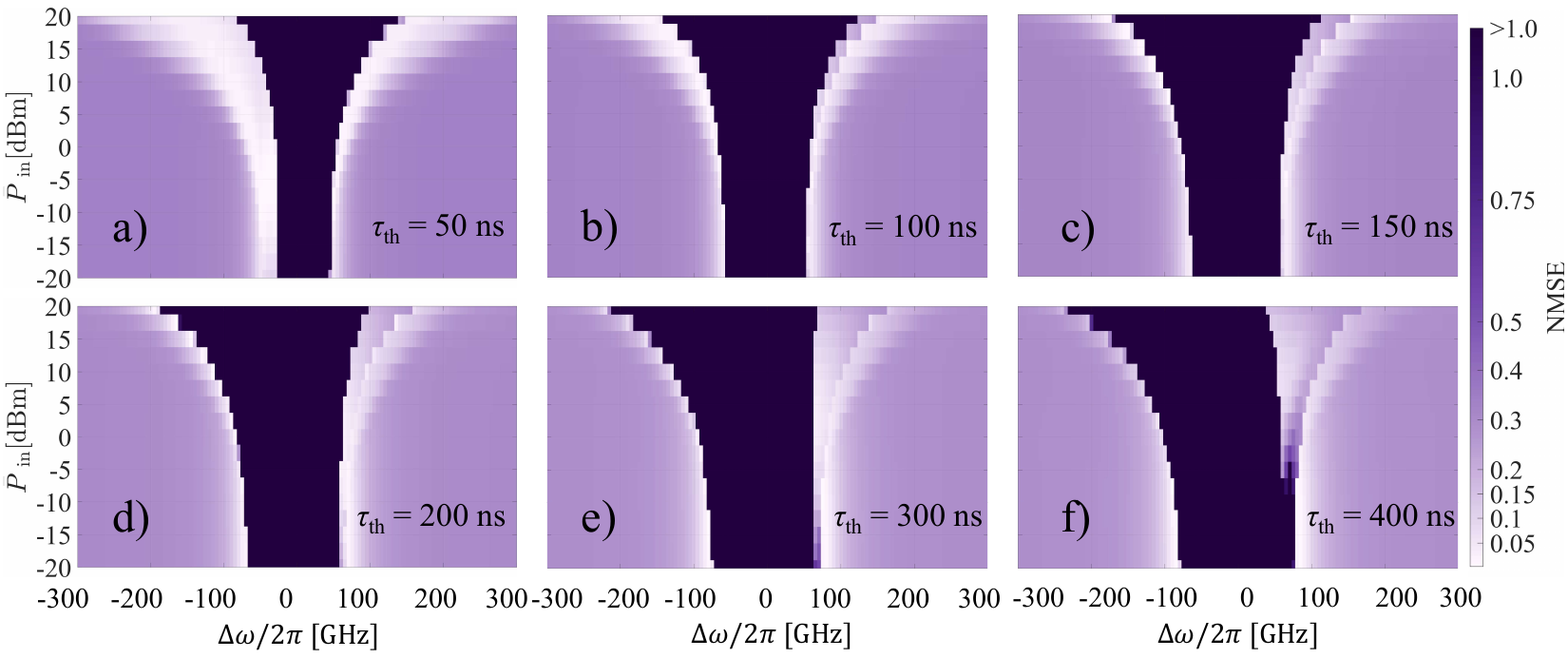}
\caption{NMSE of the simulated RC for an increasing $\tau_ \textrm{th}$ and $\tau_ \textrm{FC}$ = 10 ns, $\alpha$ = 0.8 dB/cm.}
\label{fig4}
\end{figure}

The value of $\tau_ \textrm{th}$ is related to the thermal conduction and geometry properties of the microring silicon waveguide and the cladding. It is possible to tailor $\tau_ \textrm{th}$ by altering the thickness of the cladding or its material. It is also possible to control $\tau_ \textrm{th}$ by etching trenches around the microring \cite{Johnson:06, PhysRevA.87.053805}. In Fig. \ref{fig4}, we vary $\tau_ \textrm{th}$ while fixing the values of $\tau_ \textrm{fc}$ and $\alpha$. The NMSE of the testing set prediction is obtained for the aforementioned ranges of average input power and frequency detuning. The minimum NMSE obtained for each value of $\tau_ \textrm{th}$ is shown in Table \ref{tab2}.

When increasing $\tau_ \textrm{th}$, the TO effect inside the cavity becomes dominant over the FCD effect. Hence, as mentioned in section \ref{section2}, the TO effect causes a red shift of the frequency resonance of the MRR, which is more evident in moderately higher powers (above 0 dBm) \cite{Soltani2009NovelIS}. This red shift appears to be reflected in the behaviour of region C with respect to $\omega_p$. Region C is shifted to negative detunings (redshift of $\omega_p$) also in values of $\overline{P}_{\textrm {in}}$ above 0 dBm. On the other hand, as $\tau_ \textrm{th}$ increases, region B gets narrower, and the minimum NMSE obtained is also higher. In general, increasing $\tau_ \textrm{th}$ shows to be detrimental to the performance of the RC, resulting in a limited tolerance for practical demonstration. The results of Table \ref{tab2} seem to indicate that the longer $\tau_ \textrm{th}$ the higher the required power to achieve the minimum NMSE of the results of Fig. \ref{fig4} (a-d). Nevertheless, as region C shifts towards negative detunings, the location of minimum NMSE shifts to positive detunings where now regions A and B are located.
\newpage

\begin{table}[h!]
    \centering
    \begin{tabular}[c]{|c|c|c|c|} 
         \hline
         NMSE & $\tau_ \textrm{th}$ [ns] & $\Delta\omega/2\pi$ [GHz] & $\overline{P}_{\textrm {in}}$ [dBm]\\
         \hline
         0.0178 $\pm$ 0.0018  &  50    & -30 &   -5.0 \\
        \hline
         0.0283 $\pm$ 0.0026  &  100    & -65 &   -2.5 \\   
         \hline
         0.0412 $\pm$ 0.0030  &  150    & -95 &   7.5 \\
         \hline
         0.0611 $\pm$ 0.0033  &  200    & -145 &   15 \\         
         \hline
         0.0736 $\pm$ 0.0044  &  300    & 75 &   -7.5 \\
         \hline
         0.0748 $\pm$ 0.0044  &  400    & 80 &   -7.5 \\     
         \hline
    \end{tabular}\par
    \caption{Minimum NMSE of testing set prediction for selected values of $\tau_ \textrm{th}$ ($\tau_ \textrm{FC}$ = 10 ns, $\alpha$ = 0.8 dB/cm).}
    \label{tab2}
\end{table}

\subsection{Impact of the free-carrier relaxation time}
\begin{figure}[h!]
\centering\includegraphics[width=13.3cm, trim={0 8.5cm 0 3.5cm}]{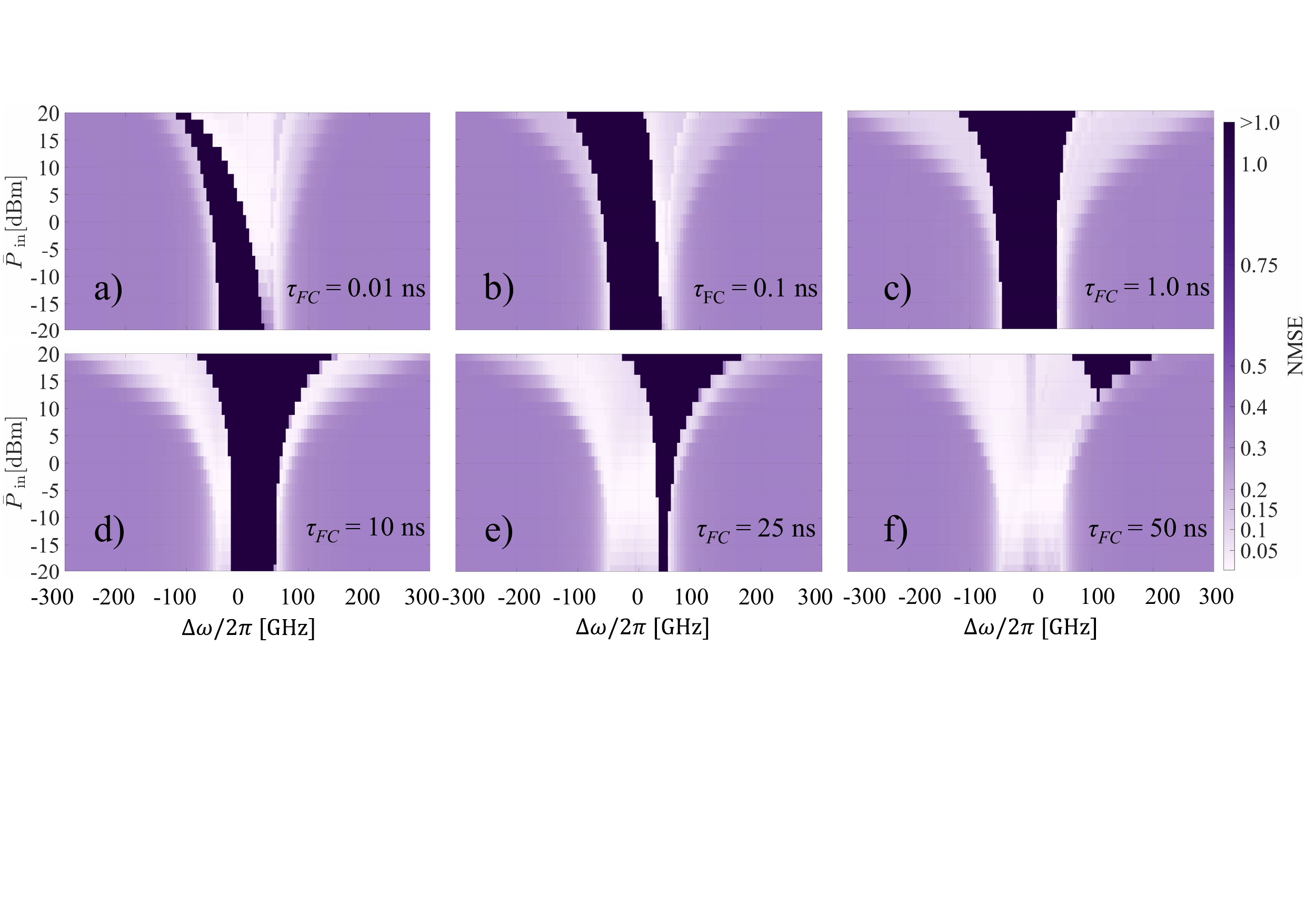}
\caption{NMSE of the simulated RC for different values of $\tau_ \textrm{FC}$ with $\tau_ \textrm{th}$ = 50 ns, $\alpha$ = 0.8 dB/cm.}
\label{fig5}
\end{figure}

The carrier's lifetime within the diffusion effective area of a microring cavity is determined by recombination processes such as the Shockley-Read-Hall (SRH), radiative and Auger recombinations. $\tau_ \textrm{FC}$ depends on the density of holes in the valence band and electrons in the conduction band \cite{554806}. In practice, the value of $\tau_ \textrm{FC}$ is usually approximated taking into account just the SRH recombination rate if the deviation of carrier density from thermal equilibrium is much larger than the density of traps \cite{Borghi:21}. $\tau_ \textrm{FC}$ can also be adjusted by improving the quality of the silicon-silica interfaces \cite{PhysRevA.87.053805}. Using the same methodology as in the previous subsection, we determine the RC performance when varying $\tau_ \textrm{FC}$ while fixing the values of $\tau_ \textrm{th}$ and $\alpha$ (Fig. \ref{fig5}).

We vary $\tau_ \textrm{FC}$ from the order of picoseconds to the order of tens of nanoseconds, so that we can analyze the impact of both reducing and increasing $\tau_ \textrm{FC}$ to have a deeper understaning of how the ratio $\tau_ \textrm{FC}$/$\tau_ \textrm{th}$ affects the RC. This ratio has a significant influence on the nonlinearities behaviour as discussed in further details later in this work. The minimum NMSE values obtained under the previous conditions are shown in Table \ref{tab3}.
For a low value of $\tau_ \textrm{FC}$ (Fig. \ref{fig5} a and b), the TO effect is dominant and, similarly to the results obtained in Fig. \ref{fig4}, region C is shifted towards negative $\Delta\omega$. However, the extension over the $\Delta\omega/2\pi$ axis seems to be reduced with respect to Fig. \ref{fig4} a) when $\tau_ \textrm{FC}$ is reduced. As $\tau_ \textrm{FC}$ increases to the order of tens of nanoseconds, the FCD effect becomes more dominant and region C shifts towards positive $\Delta\omega$, which means a blueshift of $\omega_p$. This behaviour of region C is similar to the results of the previous subsection but in the opposite direction of the spectrum. This is to be expected since TO and FCD effects cause a detuning of $\omega_0$ in opposite directions. Unlike the previous case though, when region C shifts towards either negative or positive $\Delta\omega$s its width gets narrower, which translates into a larger region B. This is of crucial importance as it opens a significantly larger window of $\overline{P}_{\textrm {in}}$ vs $\Delta\omega$ where RC operates with a very low NMSE and low power (particularly in Fig. \ref{fig5} e and f). This finding is also summarized by the results listed in Table \ref{tab3}.

\begin{table}[h!]
    \centering
    \begin{tabular}[c]{|c|c|c|c|} 
         \hline
         NMSE & $\tau_ \textrm{FC}$ [ns] & $\Delta\omega/2\pi$ [GHz] & $\overline{P}_{\textrm {in}}$ [dBm]\\
         \hline
         0.0184 $\pm$ 0.0008  &  0.01    & -25 &   10.0 \\
        \hline
         0.0192 $\pm$ 0.0023   &  0.1    & 30 &   0.0\\   
         \hline
         0.0228 $\pm$ 0.0019 &  1.0    & 45 &   -17.5 \\
         \hline
         0.0178 $\pm$ 0.0018 &  10    & -30 &   -5.0 \\         
         \hline
         0.0174 $\pm$ 0.0024  &  25    & -50 &   -5.0 \\
         \hline
         0.0151 $\pm$ 0.0021  &  50    & -45 &   -7.5 \\     
         \hline
    \end{tabular}\par
    \caption{Minimum NMSE of testing set prediction for selected values of $\tau_ \textrm{FC}$ ($\tau_ \textrm{th}$ = 50 ns, $\alpha$ = 0.8 dB/cm).}
    \label{tab3}
\end{table}

\subsection{Impact of the waveguide linear attenuation}
\begin{figure}[t!]
\centering\includegraphics[width=13.3cm, trim={0 6.5cm 0 2.5cm}]{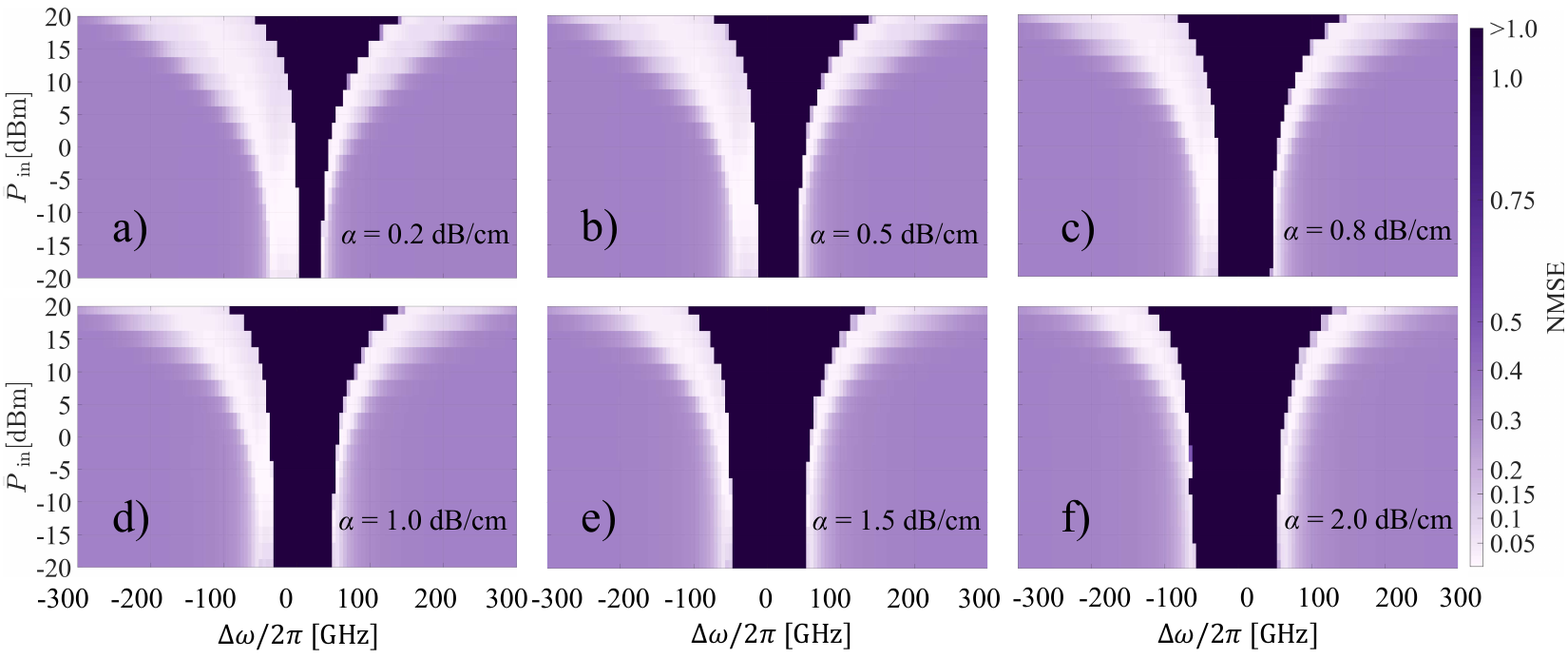}
\caption{NMSE of the simulated RC for as a function of $\alpha$, and $\tau_ \textrm{th}$ = 50 ns, $\tau_ \textrm{FC}$ = 10 ns.}
\label{fig6}
\end{figure}

The attenuation of a silicon MRR is directly related to the fabrication quality of the MRR waveguide and it has been extensively studied as high-quality factor ($Q$) silicon microring cavities have been pursued during the last decades \cite{Zeng2022-om}. However, it is also important in the context of this work to consider that modifying $\tau_ \textrm{th}$ or $\tau_ \textrm{FC}$ could lead to a collateral impact on $\alpha$, and therefore, it is important to assess the influence of $\alpha$ on the RC performance in order to fully grasp the possible performance penalties of altering the dimensions or physical properties of the MRR when tuning other parameters. Under the same frequency and average power conditions as before, we increase the value of $\alpha$ while fixing the ones of $\tau_ \textrm{th}$ and $\tau_ \textrm{FC}$ (Fig. \ref{fig6}). The minimum NMSE obtained for each increasing value of $\alpha$ is shown in Table \ref{tab4}. Contrary to the previous results, altering the waveguide attenuation does not have a great impact on the position of region C when $\overline{P}_{\textrm {in}}$ increases, although it does have an effect on its extension. A higher attenuation appears to increase the size of region C and in turn, makes region B narrower. Therefore, a relatively high-$Q$ MRR (low attenuation) is desirable in terms of performance of the RC.

\begin{table}[h!]
    \centering
    \begin{tabular}[c]{|c|c|c|c|c|} 
         \hline
         NMSE & $\alpha$ [dB/cm] & $Q$ & $\Delta\omega/2\pi$ [GHz] & $\overline{P}_{\textrm {in}}$ [dBm]\\
         \hline
         0.0164 $\pm$ 0.0020  &  0.2  & 3.5$\times10^5$ & -10 &   -7.5 \\
        \hline
         0.0169 $\pm$ 0.0020 &  0.5 & 1.4$\times10^5$   & -20 &   -5.0 \\   
         \hline
         0.0178 $\pm$ 0.0018  &  0.8 & 8.8$\times10^4$   & -30 &   -5.0 \\
         \hline
         0.0182 $\pm$ 0.0022 &  1.0 & 7.0$\times10^4$   & -45 &   -5.0 \\         
         \hline
         0.0190 $\pm$ 0.0024  &  1.5 & 4.7$\times10^4$   & -55 &   -2.5 \\
         \hline
         0.0215 $\pm$ 0.0030  &  2.0  & 3.5$\times10^4$  & 60 &   -7.5 \\     
         \hline
    \end{tabular}\par
    \caption{Minimum NMSE of testing set prediction for selected values of $\alpha$  ($\tau_ \textrm{FC}$ = 10 ns, $\tau_ \textrm{th}$ = 50 ns).}
    \label{tab4}
\end{table}

\subsection{Decreasing the number of virtual nodes}
\begin{figure}[t!]
\centering\includegraphics[width=13.3cm, trim={0 6.5cm 0 4.0cm}]{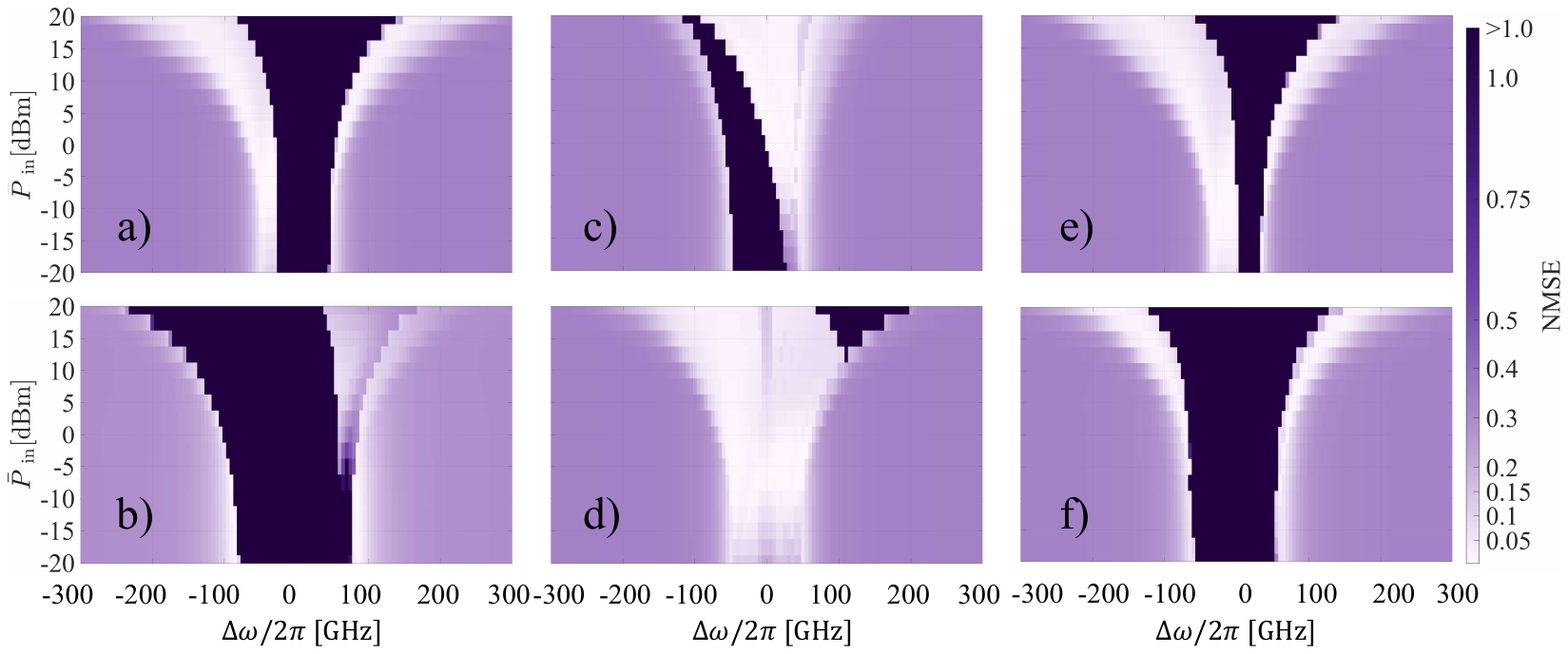}
\caption{NMSE of the simulated RC with $N =25$ for: a) $\tau_ \textrm{th}$ = 50 ns, b) $\tau_ \textrm{th}$ = 400 ns, c) $\tau_ \textrm{FC}$ = 0.01 ns, d) $\tau_ \textrm{FC}$ = 50 ns, e) $\alpha$ = 0.2 dB/cm, and f) $\alpha$ = 2.0 dB/cm.}
\label{fig7}
\end{figure}

\begin{figure}[b!]
\centering\includegraphics[width=13.3cm, trim={0 6.5cm 0 4.0cm}]{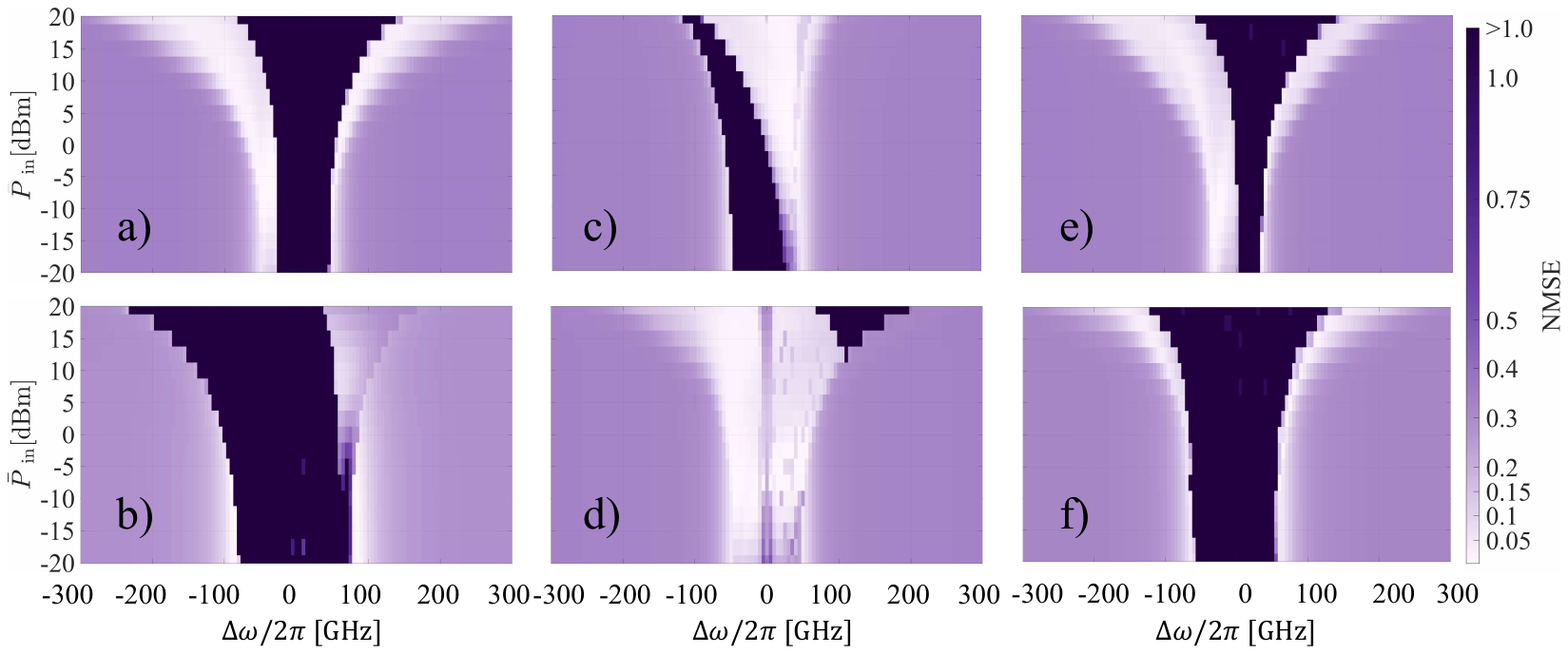}
\caption{NMSE of the simulated RC with $N =10$ for: a) $\tau_ \textrm{th}$ = 50 ns, b) $\tau_ \textrm{th}$ = 400 ns, c) $\tau_ \textrm{FC}$ = 0.01 ns, d) $\tau_ \textrm{FC}$ = 50 ns, e) $\alpha$ = 0.2 dB/cm, and f) $\alpha$ = 2.0 dB/cm.}
\label{fig8}
\end{figure}

In order to make a fair comparison with the study on the MRR-based RC with external feedback done in \cite{Donati:22}, we decrease the number of virtual nodes to 25 and simulate the RC for a sample set of configurations from the previous subsections (Fig. \ref{fig7}). This set corresponds to the minimum and maximum values of $\tau_ \textrm{th}$, $\tau_ \textrm{FC}$ and $\alpha$ simulated in the previous subsections. There is a slight increase in the obtained minimum NMSE as is expected from a neural network with fewer virtual nodes (Table \ref{tab5}). Likewise, when reducing $N$ to 10 virtual nodes, further slight degradation of the performance is observed. For this case (Fig. \ref{fig8}), the reduction of the size over the parameter space of region B is more noticeable than in the case of $N = 25$. The minimum NMSE for $N = 10$ is shown in Table \ref{tab6}.

\begin{table}[h!]
    \centering
    \begin{tabular}[c]{|c|c|c|c|c|c|} 
         \hline
         NMSE & $\tau_ \textrm{FC}$ [ns] & $\tau_ \textrm{th}$ [ns] & $\alpha$ [dB/cm] &$\Delta\omega/2\pi$ [GHz] & $\overline{P}_{\textrm {in}}$ [dBm]\\
         \hline
         0.0197 $\pm$ 0.0009  &  0.01 & 50  & 0.8 & -10 & 7.5 \\
        \hline
         0.0169 $\pm$ 0.0023  &  50  & 50  & 0.8 & -45 &   -5.0 \\
         \hline
         0.0185 $\pm$ 0.0021  &  10  & 50 & 0.8 & -40 &  -5.0 \\
         \hline
         0.0758 $\pm$ 0.0044  &  10  & 400 & 0.8 & 80 &  -7.5 \\
         \hline
         0.0173 $\pm$ 0.0025 &  10 & 50 & 0.2 & -35 &   -12.5 \\
         \hline
         0.0250 $\pm$ 0.0033   & 10 & 50 & 2.0 & 60 &   -7.5 \\     
         \hline
     
    \end{tabular}\par
    \caption{Minimum reached NMSE and corresponding parameters for the testing set prediction with $N$ = 25.}
    \label{tab5}
\end{table}

\newpage

The results indicate that even when the minimum error obtained increases, there is still no dependence of the defined $\overline{P}_{\textrm {in}}$ vs $\Delta\omega$ regions (A, B and C) on $N$, as the obtained trends are very similar between the different values of $N$. It is possible to argue that the gain in terms of minimum NMSE is relatively small when increasing $N$ up to 50. This indicates that the dimensionality of the RC is mainly given by the MRR nonlinear dynamics, and, within the considered values, $N$ does not impact the behaviour of the defined regions as much as $\tau_ \textrm{th}$,  $\tau_ \textrm{FC}$ and $\alpha$. 

\begin{table}[h!]
    \centering
    \begin{tabular}[c]{|c|c|c|c|c|c|} 
         \hline
         NMSE & $\tau_ \textrm{FC}$ [ns] & $\tau_ \textrm{th}$ [ns] & $\alpha$ [dB/cm] &$\Delta\omega/2\pi$ [GHz] & $\overline{P}_{\textrm {in}}$ [dBm]\\
         \hline
         0.0249 $\pm$ 0.0021  &  0.01 & 50  & 0.8 & 35 & 2.5 \\
        \hline
         0.0300 $\pm$ 0.0018  &  50  & 50  & 0.8 & -45 &   -7.5 \\
         \hline
         0.0224 $\pm$ 0.0019  &  10  & 50 & 0.8 & 50 &  -7.5 \\
         \hline
         0.0812 $\pm$ 0.0038  &  10  & 400 & 0.8 & 80 &  -7.5 \\
         \hline
         0.0272 $\pm$ 0.0024 &  10 & 50 & 0.2 & 35 &   -15.0 \\
         \hline
         0.0385 $\pm$ 0.0033   & 10 & 50 & 2.0 & 70 &   2.5 \\     
         \hline
     
    \end{tabular}\par
    \caption{Minimum reached NMSE and corresponding parameters for the testing set prediction with $N$ = 10.}
    \label{tab6}
\end{table}

\subsection{Characteristic RC response of the $\overline{P}_{\textrm {in}}$ vs $\Delta\omega/2\pi$ regions of NMSE}
\vspace{-1.0 cm}
\begin{figure}[hb!]
\centering\includegraphics[width=13.3cm, trim={0 2.0cm 0 8.5 cm}]{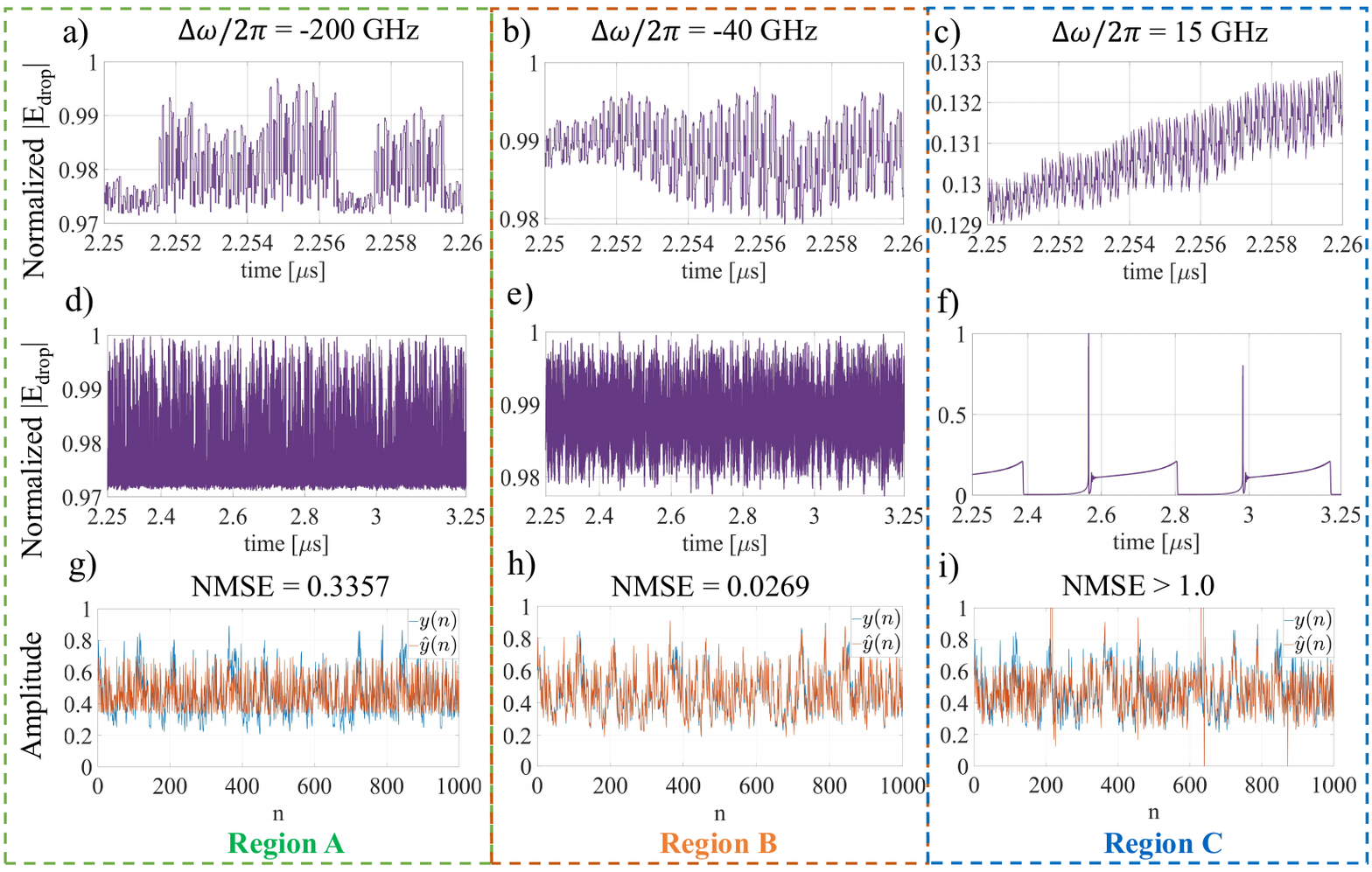}
\caption{a-c) Sampled waveforms at drop port corresponding to 10 bits of each of the three defined $\overline{P}_{\textrm {in}}$ vs $\Delta\omega/2\pi$ parameter space regions of the simulated RC when solving the NARMA-10 task with $\overline{P}_{\textrm {in}}$ = 0 dBm, N = 50, $\tau_ \textrm{th}$ = 50 ns,  $\tau_ \textrm{FC}$ = 10 ns, $\alpha$ = 0.8 dB/cm. a and d-f) 1.0 $\mu$s Extended sampled waveforms under the same conditions. g-i) Samples of the target (blue) and predicted (orange) testing sets of a NARMA-10 sequence.}
\label{fig9}
\end{figure}

In Fig. \ref{fig9} (a-f), we display an instance of the waveform of the MRR response at the drop port for each of the defined regions shown in Fig. \ref{fig2}. To switch between the regions, we fix the value of $\overline{P}_{\textrm {in}}$ to 0 dBm, while varying $\Delta\omega/2\pi$. Fig. \ref{fig9}(a-c) correspond to the response of the same 10 bits (10 ns) of the testing sequence set previously shown in \cref{fig:subim1,fig:subim2}. By zooming out this sequence to capture 1000 bits (1 \textrm{$\mu$}s), we obtain Fig. \ref{fig9}(d-f). The response of the RC in Fig. \ref{fig9}(a,d) (region A) shows a delayed linear transformation of the input in which the signal does not differ much from the input sequence. Due to  the lack of nonlinearity, the prediction fails to resemble the original NARMA-10 sequence, resulting in a relatively high NMSE = 0.3357 as the system approaches the response of a photodiode-based RC as explained in subsection \ref{sub5.1} for region A. Next, the waveform examples of region B (Fig. \ref{fig9}(b,e)) show a nonlinear transformation of the input sequence under stable conditions (avoiding SP) which gives enough dimensionality to the RC to achieve a low error (NMSE = 0.0269). Lastly, Region C, which can be observed in Fig. \ref{fig9}(c,f) shows SP oscillatory behaviour with various discontinuities or 'spikes' and fast transitions to values near zero amplitude (more visible in Fig. \ref{fig9}(f)). This behaviour of the SP affects the consistency of the RC response as in these time intervals RC is incapable of learning a consistent response to similar inputs in order to be tested with unknown data sets.

\subsection{RC linear vs nonlinear regimes}
\begin{figure}[hb!]
\centering\includegraphics[width=13cm, trim={0 7.0cm 0 3.0cm}]{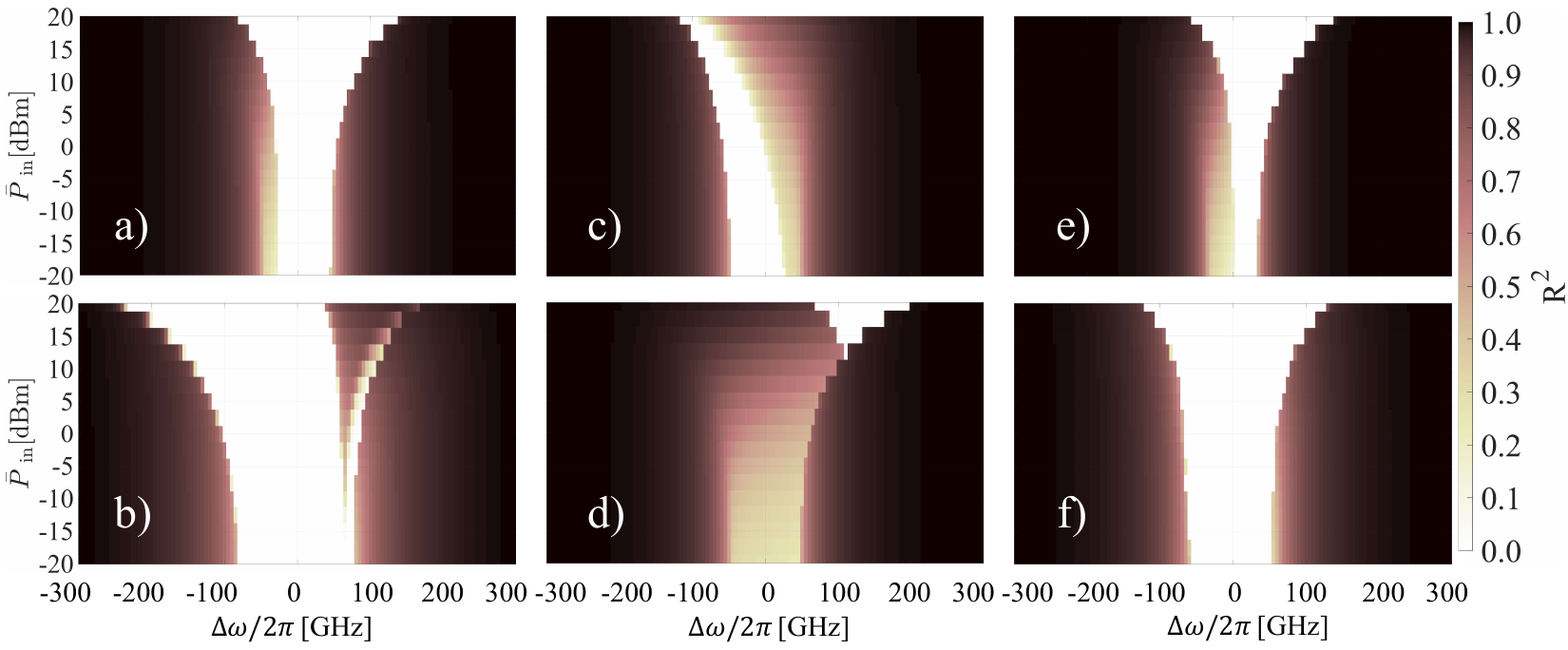}
\caption{$R^2$ between $E_ \textrm{drop}$ and $E_ \textrm{in}$ for a) $\tau_ \textrm{th}$ = 50 ns, b) $\tau_ \textrm{th}$ = 400 ns, c) $\tau_ \textrm{FC}$ = 0.01 ns, d) $\tau_ \textrm{FC}$ = 50 ns, e) $\alpha$ = 0.2 dB/cm, and f) $\alpha$ = 2.0 dB/cm.}
\label{fig10}
\end{figure}

The previous subsections qualitatively indicate the amount of nonlinear transformation the input goes through in each region. A way to quantitatively evaluate this is by determining the coefficient of determination, $R^2$, between $E_\textrm{drop}$ and $E_\textrm{in}$ so that we can quantify, on a range $[0.0 - 1.0]$, how much the response of the RC (before photodetection) can be accounted as a linear transformation of the input. In other words, an $R^2$ of 1.0 would indicate that the RC approximates a linear regime. We show the results for 6 instances of the configurations used in \cref{fig2,fig3,fig4} (2 per subsection). As demonstrated in Fig. \ref{fig10}. Each of the 3 defined regions of the previous subsections matches specific levels of $R^2$. First, it confirms that the cavity gradually gets close to a linear regime as the $R^2$ tends to 1.0 as we increase the detuning from $\omega_0$ (Region A). Then, the best performance is obtained in a mixed state between a linear and nonlinear regime where $R^2$ oscillates between $\sim$0.2 and $\sim$0.9. The location of this mixed state matches that of region B. Lastly, an $R^2$ of 0 is obtained in the same area that corresponds to SP (region C), which indicates that there is no direct relation between the response of the RC and the input. Similar findings regarding the importance of the transition between linear and nonlinear regimes to enhance TDRC performance were previously found in \cite{10.3389/fphy.2019.00138} for a Mach Zehnder Modulator used as a nonlinear node of photonic TDRC. However, in their implementation, the TDRC performs a different time-series prediction task.

\subsection{Dependence of the RC dynamics on $\Delta$T and $\Delta$N}

\begin{figure}[h!]
\centering\includegraphics[width=13.3cm, trim={0 3.0cm 0 2.25cm}]{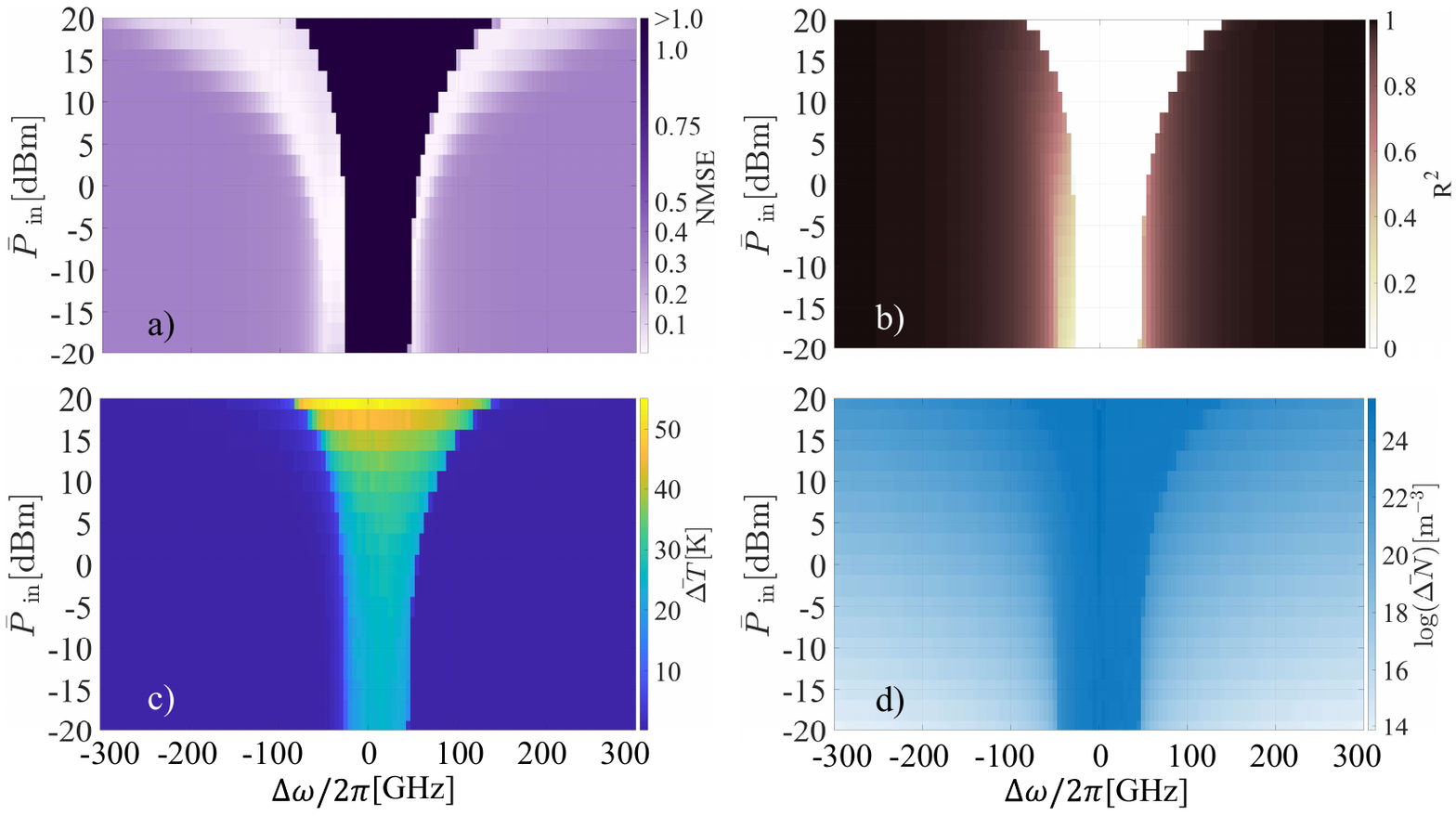}
\caption {Comparison of the $\overline{P}_{\textrm {in}}$ vs $\Delta\omega/2\pi$ heatmaps for a) NMSE of the testing set. b) $R^2$ between $E_ \textrm{drop}$ and $E_ \textrm{in}$ c) Average $\Delta T$ of the testing set computing. d) Average $\Delta N$ of the testing set computing. $\tau_ \textrm{th}$ = 50 ns,  $\tau_ \textrm{FC}$ = 10 ns, $\alpha$ = 0.8 dB/cm.}
\label{fig11}
\end{figure}

Finally, we also assess the relation between the defined regions (A, B and C) over the $\overline{P}_{\textrm {in}}$ vs $\Delta\omega/2\pi$ parameter space, with $\Delta T$ and $\Delta N$. To achieve this, we plot the average of $\Delta T$ and $\Delta N$ for the Runge-Kutta solution of \cref{eq1,eq2,eq3} of the whole testing set. By comparing their $\overline{P}_{\textrm {in}}$ vs $\Delta\omega/2\pi$ heatmaps with the ones related to the NMSE and $R^2$, as shown in Fig. \ref{fig11}, we can see a clear relation between the increase in free-carrier concentration and temperature, and the rise of SP with its previously stated effects on the RC response. Very low average $\Delta N$ or $\Delta T$ are not good either for the performance of the RC, as this shows to be highly correlated with the MRR approximating a linear regime.
To achieve optimum performance, relatively small increases in temperature $\sim$(2.5 to 15 K) and a $\Delta N$ between $\sim$10$^{18}$ and $\sim$10$^{22}$ appear to be the key requirements. However, the susceptibility of the MRR cavity to enable SP for a given $\Delta\omega$ and $\overline{P}_{\textrm {in}}$ also depends strongly on the ratio between the relaxation lifetimes of the nonlinear effects as discussed in section \ref{section6}.

\section{Discussion}\label{section6}
In our study, the free-carrier and TO nonlinearities relaxation times ($\tau_ \textrm{FC}$, $\tau_ \textrm{th}$) and the waveguide attenuation $\alpha$ have been varied while fixing the other parameters in order to understand the individual contribution of each of them to the overall performance of the RC. However, in practice, each of them is highly correlated with the others and the potential changes that can be made to the MRR to optimize one of them will most probably affect the others. In the case of $\tau_ \textrm{FC}$ and $\tau_ \textrm{th}$ it is also essential to consider the ratio between them, $\tau_ \textrm{FC}$/$\tau_ \textrm{th}$. As shown in \cite{PhysRevA.87.053805, PhysRevA.86.063808}, this ratio has a severe influence on the existence of SP, depending in turn on $\Delta\omega$ and $\overline{P}_{\textrm {in}}$. In \cite{PhysRevA.86.063808} the authors conclude that in order to enhance SP, a short $\tau_ \textrm{FC}$ is required with respect to $\tau_ \textrm{th}$, but not too short (it should be longer than the photon lifetime, i.e., the relaxation time of the resonance $\tau_ \textrm{r} = \frac{n_{\textrm {Si}}}{c\alpha}$). This condition is matched in the simulated parameters used to obtain the results of Fig. \ref{fig4}. By increasing $\tau_ \textrm{th}$, the ratio $\tau_ \textrm{FC}$/$\tau_ \textrm{th}$  becomes smaller and since $\tau_ \textrm{FC}$ = 10 ns is higher than $\tau_ \textrm{r}$ ($\sim$0.14 ns for $\alpha$ = 0.8 dB/m), conditions are favourable to reinforce SP and extend its region. Due to the increase of $\tau_ \textrm{th}$, the TO effect is dominant and the SP region is shifted towards negative detuning (red-shift of the resonance) \cite{PhysRevA.87.053805}.   

Hence, for a given $\alpha$, one way to diminish SP behaviour is to reduce $\tau_ \textrm{FC}$ as much as possible, so that the requirement $\tau_ \textrm{FC}$<$\tau_ \textrm{r}$ is fulfilled. A few attempts to try to reduce $\tau_ \textrm{FC}$ to sub-nanosecond magnitudes are found in the literature: A $\tau_ \textrm{FC}$ of 55 ps is achieved in \cite{Forst:07} using ion implantation in a silicon MRR, although the waveguides are penalized with additional 22 dB/cm propagation losses. Moreover, a $\tau_ \textrm{FC}$ of 15 ps is reported in \cite{Waldow:08} using an oxygen-implanted silicon on insulator (SOI) MRR for all-optical switching purposes. Lastly, in \cite{Turner-Foster:10}, a p-i-n junction is integrated into an SOI waveguide to reduce $\tau_ \textrm{FC}$ to 12.2 ps with an attenuation loss of 2 dB/cm.

Another way to keep a certain level of nonlinear dynamics without reinforcing SP would be to increase the ratio $\tau_ \textrm{FC}$/$\tau_ \textrm{th}$ \cite{PhysRevA.86.063808}. This can be done by decreasing $\tau_ \textrm{th}$ so that both relaxation times are around the same order of magnitude (Fig. \ref{fig4}(a)), but this is more challenging as $\tau_ \textrm{th}$ is very dependent on the quality of the fabrication process and the geometry of the silicon waveguide. $\tau_ \textrm{th}$ is also relatively fixed by the required width of the cladding material. So, the other way to increase $\tau_ \textrm{FC}$/$\tau_ \textrm{th}$ would be to increase $\tau_ \textrm{FC}$, and with it, the strength of free-carrier nonlinearities. Our results are coherent with these previous methods as the area of very high error (Region C) that corresponds to SP is reduced by making $\tau_ \textrm{FC}$  either too small in comparison to $\tau_ \textrm{th}$ and smaller than $\tau_ \textrm{r}$ (order of picoseconds, Fig. \ref{fig5}(a)) or by making $\tau_ \textrm{FC}$/$\tau_ \textrm{th}$ >= 0.5 (Fig. \ref{fig5}(b)). Both scenarios provide enough nonlinearity to the RC and region B is extended to lower levels of power.

Regarding the linear losses, resonator cavities with a high $Q$-factor are indispensable to achieve the necessary nonlinear dynamics for the RC to operate with low NMSE \cite{PhysRevA.86.063808}. Higher linear losses contribute to the increase in heat due to power absorption and in turn, the TO effect becomes the dominant effect and the region of SP is widened. On the contrary, low linear waveguide losses (which normally translate into relatively high $Q$ MRRs) allow the dynamics of free-carrier nonlinearities to become more relevant within the cavity. As FCD becomes more dominant, the resonance is blue-shifted and the SP region shifts to positive detunings \cite{PhysRevA.86.063808}. Our results in Fig. \ref{fig6} match the previously described physical dynamics.   

Following the aforementioned insights from studies about SP in MRR cavities, therefore, provides a way to enhance the range of $\overline{P}_{\textrm {in}}$ and $\Delta\omega/2\pi$ in which region B occurs. These conditions are consistent with our results. The existence of this interval of low-error prediction (region B in our case) also fits well with the results of \cite{Boikov_2023}, where, as we mentioned before, a power interval is also found for FCD nonlinearities in which the RC is given enough dimensionality while keeping consistency using directly coupled cavities.

As a matter of performance comparison, we state previous works about photonic TDRC schemes solving the NARMA-10 task with the same number of virtual nodes ($N$ = 50) and similar speed. An optoelectronic scheme using a Mach-Zehnder modulator achieved an NMSE = 0.168 $\pm$ 0.015 \cite{Paquot2012}. Even though, we acknowledge that the results in \cite{Paquot2012} were obtained both in simulations and experimentally. Numerically, an NMSE = 0.103 $\pm$ 0.018 was obtained in \cite{Chen:19}. In order to perform a fair comparison of our results with \cite{Donati:22}, we simulate the system using the highest and lowest values of the nonlinear effects lifetimes and waveguide loss considered in this study, for $N$ = 25 nodes. The minimum NMSE with that number of virtual nodes numerically obtained in \cite{Donati:22} is 0.204 $\pm$ 0.026. In our results, we obtain a minimum NMSE of 0.0151 $\pm$ 0.0021, which is a considerable improvement over \cite{Donati:22} and we could expand the region in which the RC achieves its best performances in comparison to our previous study \cite{castro2023impact}.

\section{Conclusion}\label{section7}

Throughout this work, we have investigated the relationship between the performance of an MRR-based TDRC and the lifetimes of the free-carrier and TO effects as well as the impact of the waveguide attenuation. We simulate different values for the parameters ($\tau_ \textrm{FC}$, $\tau_ \textrm{th}$ and $\alpha$), and define three $\overline{P}_{\textrm {in}}$ vs $\Delta\omega/2\pi$ regions in which each of them shows a very different level of error prediction. We characterize such regions by first showing qualitatively the waveform response of each region and then we quantify the differences in nonlinearity between such regions using $R^2$. We show that when the response at the drop port can be potentially represented by a linear transformation of the input sequence, the nonlinearities of the system are given mostly by the photodiode response at the detection stage. When the $R^2$ between the output and input of the RC is very low, the system might run into the risk of obtaining a response too inconsistent to accurately calculate the weights during the training process due to the discontinuities and near-to-zero response of the RC caused by SP nonlinearities. There is an interval of average input power and angular frequency detuning in which enough dimensionality is given to the RC without becoming unstable due to SP.  This work provides a further understanding of the physical conditions that are optimum to reduce SP while keeping a high dimensionality. In the areas that fulfil such conditions, our RC achieves low error at low power when solving a time-series prediction. Our results obtain an NMSE lower than other works that propose a similar $N$ and speed. We also show that decreasing $N$ does not have a great impact on the physical dynamics of this RC setup.

\medskip
\begin{backmatter}
\bmsection{Funding}
Villum Fonden (OPTIC-AI grant no.VIL29334), and by the Swedish Research Council (VR) project BRAIN (2022-04798).
\bmsection{Disclosures}
The authors declare no conflicts of interest.

\bmsection{Data Availability}
Data underlying the results presented in this paper are not publicly available at this time but may be obtained from the authors upon reasonable request.

\end{backmatter}

\bibliography{Opt_Express_509437_Bernard_Giron}

\end{document}